\documentclass[11pt,twocolumn,prb,superscriptaddress,reprint,aps]{revtex4-1}

\usepackage{graphicx,amsmath,amssymb,color}
\usepackage{tabularx, ctable}
\usepackage{threeparttable}
\usepackage{ulem}
\usepackage{bm}
\usepackage{siunitx}
\usepackage{gensymb}
\usepackage{makecell}

\usepackage[breaklinks=true,colorlinks,allcolors=blue]{hyperref}

\newcommand{\be}{\begin{eqnarray}}
\newcommand{\ee}{\end{eqnarray}}

\makeatletter
\let\auto@bib@innerbib\@empty
\makeatother

\newcounter{suppnote}
\newcommand{\suppnote}[1]{%
  \refstepcounter{suppnote}%
  \par\vspace{1em}
  {\centering
   \fontsize{11}{13}\selectfont
   \bfseries
   Supplementary Note \thesuppnote.%
   \hspace{0.5em}%
   \MakeUppercase{#1}%
   \par}
  \vspace{0.8em}
}

\let\MergedTitleCommand\title
\let\MergedAuthorCommand\author
\let\MergedAffiliationCommand\affiliation
\let\MergedMakeTitleCommand\maketitle

\begin{document}

\title{Correlated Insulator Moiré Bolometer}

\author{L. Elesin$^{+}$}
\affiliation{Department of Materials Science and Engineering, National University of Singapore, 117575, Singapore}

\author{A. L. Shilov$^{+}$}
\affiliation{Department of Materials Science and Engineering, National University of Singapore, 117575, Singapore}

\author{M. Kravtsov$^{+}$}
\affiliation{Department of Materials Science and Engineering, National University of Singapore, 117575, Singapore}

\author{X. Zhou}
\affiliation{Department of Materials Science and Engineering, National University of Singapore, 117575, Singapore}

\author{M. Lukianov}
\affiliation{Center for Neurophysics and Neuromorphic Technologies, Moscow, 127495, Russia}

\author{A. Kuksov}
\affiliation{Center for Neurophysics and Neuromorphic Technologies, Moscow, 127495, Russia}

\author{S. Jana}
\affiliation{Department of Materials Science and Engineering, National University of Singapore, 117575, Singapore}

\author{I. Iorsh}
\affiliation{Queen's University, Kingston, K7L 3N6, Canada}

\author{R. Izmaylov}
\affiliation{Moscow Pedagogical State University, Moscow, 119991, Russia}

\author{K. Shein}
\affiliation{Moscow Pedagogical State University, Moscow, 119991, Russia}

\author{I. Gayduchenko}
\affiliation{Moscow Pedagogical State University, Moscow, 119991, Russia}

\author{T. Taniguchi}
\affiliation{International Center for Materials Nanoarchitectonics, National Institute for Materials Science, Tsukuba, 305-0044, Japan}

\author{K. Watanabe}
\affiliation{Research Center for Functional Materials, National Institute for Materials Science, Tsukuba, 305-0044, Japan}

\author{K. S. Novoselov}
\affiliation{Institute for Functional Intelligent
Materials, National University of Singapore, Singapore, 117575, Singapore}

\author{G. N. Goltsman}
\affiliation{Moscow Pedagogical State University, Moscow, 119991, Russia}

\author{A. Principi}
\affiliation{School of Physics and Astronomy, University of Manchester, Manchester, M13 9PL, United Kingdom}

\author{D. A. Bandurin$^{*}$}
\affiliation{Department of Materials Science and Engineering, National University of Singapore, 117575, Singapore}
\affiliation{Institute for Functional Intelligent
Materials, National University of Singapore, Singapore, 117575, Singapore}

\maketitle

\begin{center}
$^{*}$ Correspondence to: dab@nus.edu.sg

$^{+}$ These authors contributed equally.
\end{center}

\section*{Abstract}
Light incident on an insulator is generally not expected to turn it into a metal without invoking intense ultrafast excitation that leads to transient structural transitions. Here we show that magic-angle twisted bilayer graphene tuned to half filling of the moiré band provides a notable exception to this expectation. We find that weak long-wavelength photons, with energies comparable to the flat-band width, selectively heat the low-heat-capacity electronic subsystem, thereby suppressing the correlated gap. This produces a giant resistance change governed not by a persistent photocarrier population, but by the extreme sensitivity of a many-body correlated gap to weak electronic heating. The resulting photon-driven insulator-to-metal transition produces a broadband low-noise photoresponse with voltage responsivity exceeding millivolts per nW of absorbed power. The mechanism is dual to superconducting hot-electron response: radiation-heated electrons suppress a many-body order, but in reverse the correlated insulator melts into a metal, providing robustness to magnetic fields of several tesla and a sharp insulator-to-metal resistive contrast. Our results establish correlated flat-band systems as a platform for ultrasensitive detection of faint long-wavelength radiation.

\section*{Introduction}

Light incident on a material can alter its electrical resistance -- a class of phenomena termed photoresistance~\cite{Fox2010OpticalProperties}. In conventional semiconductors, absorbed photons promote electrons across a bandgap, generating free carriers that enhance conductivity. In metals, by contrast, photoresistance is typically negligible: although absorption may heat the electron gas, the large Fermi energy and high carrier density render this perturbation largely irrelevant for transport. Reducing dimensionality changes this picture qualitatively: resonant excitations in quantum dots, hot-carrier and thermoelectric effects in low-dimensional conductors, and other mechanisms can strongly amplify the impact of optical excitation on transport, giving rise to pronounced photoconductive responses and enabling new device functionalities~\cite{antononv,PhotoconductivityCNT,Xia2009Ultrafast,Gabor,TBGphotoconductivity,shilov2025TBG}. Yet a unifying feature of these standard effects is their single-particle character: photons couple to individual electrons, redistributing them in energy or momentum and thereby modifying resistance through changes in carrier density or scattering.

\begin{figure*}[ht!]
  \centering\includegraphics[width=1\linewidth]{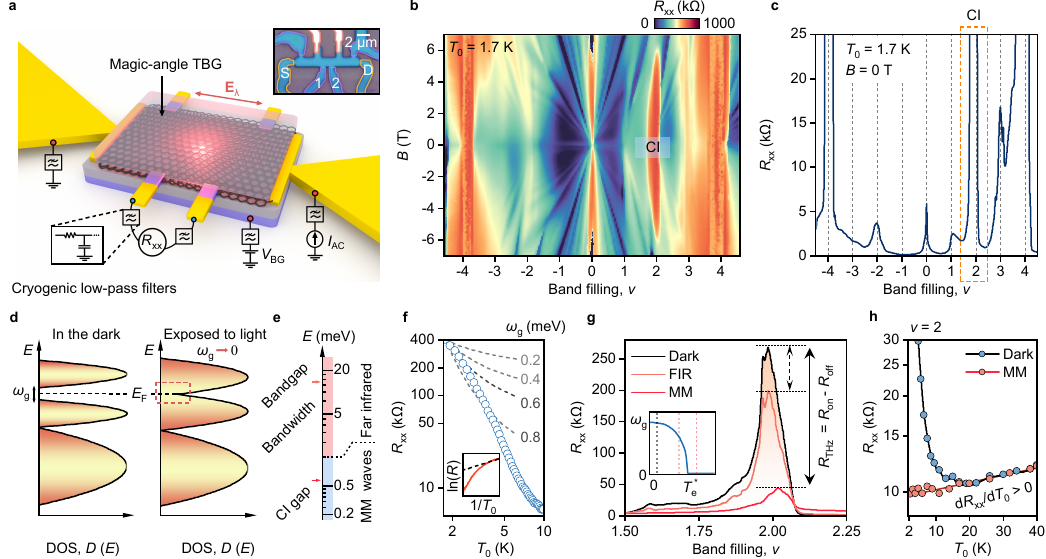}
    \caption{\textbf{Correlated Insulator Moiré Bolometer.}
\textbf{a}, Schematic of the magic-angle twisted bilayer graphene (TBG) device. The red spot indicates hot electrons generated in the TBG channel under THz irradiation with electric field $\mathbf{E}_{\lambda}$, where $\lambda$ is the radiation wavelength. The longitudinal resistance $R_\mathrm{xx}$ was measured in a four-probe configuration using an AC excitation current $I_\mathrm{AC}$, while the carrier density was tuned by the graphite back-gate voltage $V_{\mathrm{BG}}$. Inset: optical image showing the source (S), drain (D), and voltage probes.
\textbf{b}, Landau-fan diagram of $R_\mathrm{xx}$ measured at the cryostat base temperature $T_0=1.7$ K, showing a pronounced correlated insulator (CI) state.
\textbf{c}, $R_\mathrm{xx}$ versus band filling; the dashed rectangle marks the CI state.
\textbf{d}, Illustration of the density of states (DOS) in the dark and under irradiation. At half-filling, the Fermi level $E_{\mathrm{F}}$ lies within the correlated gap $\omega_{\mathrm{g}}$. Under sub-THz irradiation, the gap closes, $\omega_{\mathrm{g}}\rightarrow0$, producing a finite DOS at $E_{\mathrm{F}}$ highlighted by the red dashed box.
\textbf{e}, Energy scale in the system.
\textbf{f}, Collapse of the CI state with increasing temperature. Gray dashed lines show Arrhenius fits yielding the corresponding gaps $\omega_{\mathrm{g}}$. Inset: the same data in Arrhenius coordinates, showing behavior consistent with Ref.~\cite{shavit2023strain}; the black dashed line is the limiting Arrhenius fit as $T\rightarrow T_0=1.7$ K. Data were measured at $500~\mathrm{pA}$ to minimize current-induced overheating.
\textbf{g}, Comparison of $R_\mathrm{xx}$ measured in the dark ($R_{\mathrm{off}}$) and under far-infrared (FIR) or millimetre-wave (MM) irradiation ($R_{\mathrm{on}}$). The photoresistance is defined as $R_{\mathrm{THz}}(\nu)=R_{\mathrm{on}}(\nu)-R_{\mathrm{off}}(\nu)$. Inset: qualitative dependence of $\omega_{\mathrm{g}}$ on electronic temperature $T_{\mathrm{e}}$\cite{shavit2023strain}; the gap vanishes with increasing temperature ($T_\mathrm{e}^{*}$ denotes the critical electron temperature), and dashed lines indicate the radiation-induced increase in $T_{\mathrm{e}}$.
\textbf{h}, Temperature dependence of $R_\mathrm{xx}$ with and without MM-wave illumination at $\nu=2$. Under illumination, the device shows metallic behavior, $\mathrm{d}R_\mathrm{xx}\mathrm{/d}T_0>0$. This curve was acquired during a different cooldown; therefore, the absolute resistance values differ from those shown in Fig.~1g.
}
	\label{Fig1}
\end{figure*}

\begin{figure*}[ht!]
  \centering
  \includegraphics[width=\linewidth]{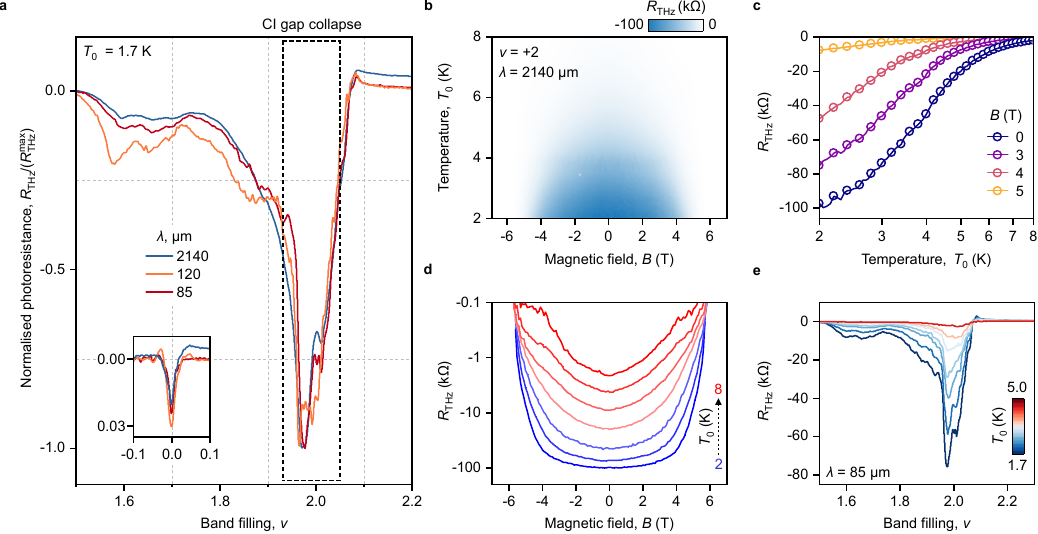}
  \caption{\textbf{Broadband photoresistance of the CI state.} 
  \textbf{a}, Normalised photoresistance $R_{\mathrm{THz}}/R^\mathrm{max}_{\mathrm{THz}}$ as a function of filling factor $\nu$ near the CI state ($\nu = 2$) for different wavelengths. The dashed box highlights the region of the most pronounced negative photoresistance near $\nu=2$. Inset: normalised photoresistance in the vicinity of the charge-neutrality point (CNP).  
  \textbf{b}, Phase diagram of the photoresistance as a function of external magnetic field and temperature. 
  \textbf{c}, Temperature-dependent photoresistance traces at selected magnetic fields. 
  \textbf{d}, Magnetic-field dependence of photoresistance at the CI state for selected temperatures. 
  \textbf{e}, Band-filling dependence of photoresistance under illumination with $\lambda = 85~\mu\mathrm{m}$ radiation at various temperatures.}
  \label{Fig2}
\end{figure*}

A handful of exceptions illustrate a qualitatively different regime. In ultra-clean low-dimensional conductors near the hydrodynamic regime, light-induced heating can drive a crossover from ballistic to viscous electron flow, producing a photoresponse governed by collective transport coefficients~\cite{kravtsov2025viscous}. Furthermore, plasmons -- collective self-sustained oscillations of electron density - can resonantly enhance absorption of low-dimensional conductors~\cite{Bandurin2018,Caridad2024,Bandurin2022,Frederic} and thereby can influence the photoresponse. Electromagnetic fields can also directly excite collective charge-order dynamics, as recently demonstrated by atomically resolved THz spectroscopy of charge-density-wave phase excitations in NbSe$_2$~\cite{sheng2024}. More dramatically, in superconducting thin films and nanowires, absorbed photons readily suppress the zero-resistance state, enabling ultrasensitive photodetection~\cite{johnson1996bolometric,semenov2001quantum,gol2001picosecond,di2024infrared}. This example points to a broader principle: when electrical resistance is governed by a many-body order parameter rather than single-particle band structure, photons need only destabilize that order to produce a dramatic response. Correlated two-dimensional quantum materials provide a natural platform to revisit this idea, as their transport properties are defined by emergent electronic order stabilized by interaction energies that can be remarkably small~\cite{cao2018correlated,cao2018unconventional,sharpe2019emergent,yankowitz2019tuning,serlin2020intrinsic,andrei2020graphene,he2021moire,balents2020superconductivity,nuckolls2024microscopic}. Yet despite intense fundamental interest in these materials, their optoelectronic behavior almost always~\cite{di2024infrared,krishna2025terahertz} has remained confined to single-particle physics~\cite{Deng2020,intelligent,RoshSNSPD,MoireOptoReview}. Graphene-based moiré flat-band systems are particularly compelling in this context: their fragile correlated insulating (CI) states are controlled by energy scales of only a few millielectronvolts -- often the smallest scale in the system~\cite{choi2021interaction,kim2023imaging,kim2026resolving,nuckolls2023quantum,shavit2023strain}. This raises a fundamental question: can weak, low-energy photons destabilize such an insulator and drive it metallic, producing a giant photoresistive response without invoking structural transitions or lattice dynamics?~\cite{Cavalleri2001,Rini2007,Stojchevska2014} Until now, this was thought to require the brute force of intense, above-gap femtosecond excitation that injects carriers across eV-scale Mott or charge-transfer gaps~\cite{Iwai2003,Okamoto2007,Perfetti}, thereby overlooking a regime in which weak perturbations can produce disproportionately large electronic responses.

Here we show that this expectation does not apply to magic-angle twisted bilayer graphene (MATBG) and demonstrate giant photoresponse of MATBG to low-intensity long-wavelength photons when it is tuned to half filling of the moiré band~\cite{choi2021interaction,kim2023imaging,kim2026resolving,nuckolls2023quantum,shavit2023strain}.
Far-infrared photons selectively overheat the low-heat-capacity electronic subsystem~\cite{aamir2021ultrasensitive,di2022revealing}, driving a collapse of the correlated order and producing a giant photoresistive response. By studying the dependence on band filling, temperature, and radiation power, we show that the effect originates from the destabilization of the many-body gap rather than from single-particle excitation. Beyond its implications for correlated-electron physics, this mechanism enables highly sensitive far-infrared detection with megavolt-per-watt internal responsivity and low noise. 

\section*{Results}

\begin{figure*}[ht!]
  \centering\includegraphics[width=\linewidth]{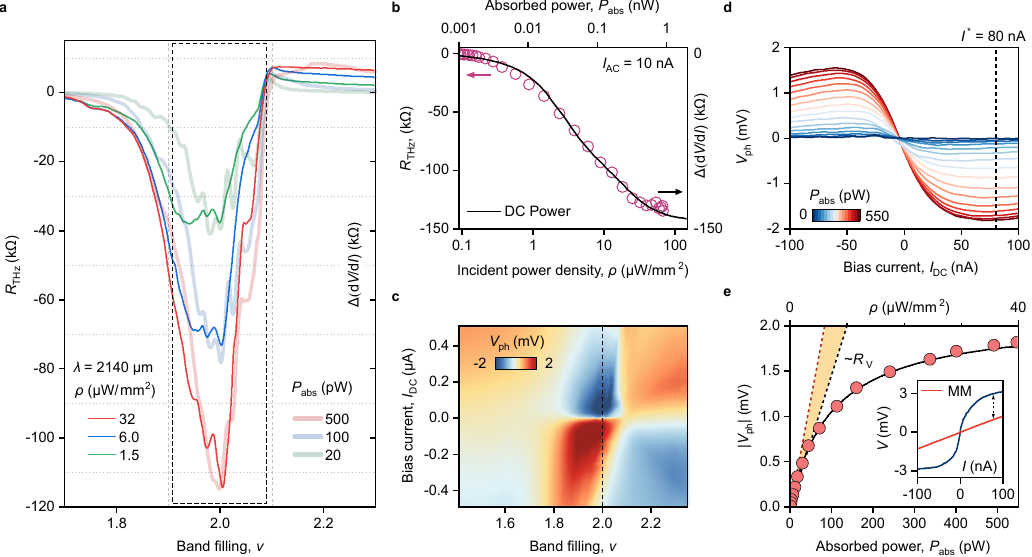}
    \caption{\textbf{Responsivity of the CI moiré bolometer.}
    Photoresponse data shown in this figure were acquired at $\lambda = 2140~\mu\mathrm{m}$.
\textbf{a}, Comparison of the THz-induced photoresistance $R_{\mathrm{THz}}(\nu)$ near the CI with the change in differential resistance $\mathrm{d}V\mathrm{/d}I(P_{\mathrm{abs}})-\mathrm{d}V\mathrm{/d}I(0)$ produced by an equivalent absorbed power $P_{\mathrm{abs}}$ generated via DC heating from a DC bias current $I_{\mathrm{DC}}$. The dashed box highlights the selected region near $\nu=2$.
\textbf{b}, Comparison between the photoresistance at $\nu = 2$ as a function of incident THz power density and the change in $\mathrm{d}V\mathrm{/d}I(P_{\mathrm{abs}})-\mathrm{d}V\mathrm{/d}I(0)$ produced by DC heating with power $P_{\mathrm{abs}}$.
\textbf{c}, Photovoltage $V_{\mathrm{ph}}$ mapped as a function of $\nu$ and DC bias $I_{\mathrm{DC}}$.
\textbf{d}, $V_{\mathrm{ph}}$ vs bias current $I_{\mathrm{DC}}$ under THz radiation illumination of different power (the incident power is recalculated to absorbed power using the calibration curve). The dashed line represents the optimal operation point  $I^*$ of the bolometer.
\textbf{e},  $|V_{\mathrm{ph}}|$ as a function of $P_{\mathrm{abs}}$ under THz illumination. The data were acquired at a fixed DC bias current of $I^* = 80\,\mathrm{nA}$. Dashed lines represent the responsivity of the device in the linear regime $R_\mathrm{V}$, and the solid line is a saturation fit $|V_\mathrm{ph}|=V_0\frac{P_\mathrm{abs}}{P_0+P_\mathrm{abs}}$ (for $V_0 = 2.1$ mV and $P_0=100$ pW). Inset: $V$--$I$ characteristics under MM-wave irradiation and in the dark: the difference in voltage at fixed current corresponds to the photovoltage. 
}
	\label{Fig3}
\end{figure*}

\textbf{Correlated Insulator Moiré Bolometer.}
To explore the bolometric capabilities of the CI state, we fabricated a device based on MATBG, encapsulated in two hexagonal boron nitride (hBN) slabs (see Methods). The device was patterned in the form of a multiterminal field-effect transistor (FET) with a global graphite back gate, enabling precise control of carrier doping of the whole device (Fig.~\ref{Fig1}a). 

Figure~\ref{Fig1}b presents the longitudinal resistance, $R_\mathrm{xx}$, mapped as a function of moiré band filling, $\nu = n/n_{\mathrm{0}}$ (where $n$ is the electrostatically doped carrier concentration and $4n_{\mathrm{0}}$ represents the full filling of superlattice minibands), and perpendicular magnetic field, $B$, together with its zero-field cross-section (Fig.~\ref{Fig1}c). The data exhibit the characteristic features of MATBG: a charge-neutrality point at $\nu = 0$, strong single-particle insulating states at $\nu = \pm 4$, and CIs at $\nu = \pm 2$, with the $\nu = +2$ state being particularly pronounced.

Figure~\ref{Fig1}f displays the temperature dependence of $R_\mathrm{xx}$ measured at $\nu=2$. The insulating behavior is remarkably steep: $R_\mathrm{xx}$ increases by more than an order of magnitude upon cooling the sample from 10~K to 1.7~K. In this temperature window, $R_{\mathrm{xx}}(T_0)$ clearly deviates from the simple Arrhenius dependence expected only for a temperature-independent gap (Fig.~\ref{Fig1}f)~\cite{shavit2023strain}. 
This behavior is consistent with previous scanning tunneling microscopy studies~\cite{choi2021interaction,kim2023imaging,kim2026resolving}, which identified that the ground state at $\nu \approx 2$ is a symmetry-broken insulating phase with an intervalley-coherence (IVC) order, characterized by an energy gap of approximately 1~meV below 1~K. This gap is gradually closing upon raising $T$ above 1.5 K and fully vanishes around $10~$K\cite{choi2021interaction,kim2023imaging,kim2026resolving,shavit2023strain}, consistent with our transport data.

\textbf{Collapse of the CI and giant photoresistance.} 
Figure~\ref{Fig1}g demonstrates the extreme sensitivity of the sample resistance at $\nu = 2$ to incident millimetre-wave (MM) and far-infrared (FIR) photons. Photon absorption heats the electronic subsystem, destabilizing the fragile CI state and reducing the sample resistance. This sensitivity arises from the following mechanism. 

External illumination increases the effective carrier temperature in the flat band (see the results of the independent Johnson noise thermometry measurements in Supplementary Figure 12). 
This, in turn, leads to the destabilization of the CI, restoring a finite density of states near the Fermi level (see Figure \ref{Fig1}d) and producing a dramatic drop in resistivity. The effect is conceptually symmetric to that of superconducting detectors. In both cases, electron heating perturbs a correlated electronic phase; however, while superconducting bolometers rely on the destruction of a zero-resistance state, our MATBG bolometer exploits photon-driven CI to metal transition. The $R_\mathrm{xx}$ dependence on $T_\mathrm{0}$ corroborates this interpretation by comparing the data taken in the dark and under MM-wave excitation (Fig.~\ref{Fig1}h). While the former signals insulating temperature dependence, the latter reveals a clear $\mathrm{d}R_\mathrm{xx}\mathrm{/d}T_\mathrm{0}>0$ trend -- an indication of a metallic state (See Supplementary Figure 17 for the full phase diagram). Since the CI gap constitutes the smallest energy scale in the system (Fig.~\ref{Fig1}e), the bolometric response extends over a broad spectral range, encompassing the technologically important MM-wave and FIR bands.

To facilitate further analysis, we define the photoresistance as
$R_{\mathrm{THz}}(\nu) = R_{\mathrm{on}}(\nu) - R_{\mathrm{off}}(\nu)$,
where $R_{\mathrm{off}}$ and $R_{\mathrm{on}}$ denote resistance values recorded in the dark and under illumination, respectively (see Methods). Figure~\ref{Fig2} summarizes the photoresponse characteristics of the MATBG bolometer. The normalised photoresistance traces $R_{\mathrm{THz}}/R^\mathrm{max}_{\mathrm{THz}}$ in Fig.~\ref{Fig2}a (where $R^\mathrm{max}_{\mathrm{THz}}$ is the maximum photoresistance at a given wavelength and power) reveal a pronounced response of the CI state at $\nu = 2$ to incident radiation over a wide spectral range, from $\lambda = 85~\mu\mathrm{m}$ to $\lambda = 2140~\mu\mathrm{m}$. In contrast, outside the CI the system exhibits a finite and positive $R_{\mathrm{THz}}$, consistent with metallic behavior (Supplementary Figure 1e shows quantum oscillations in the resistance as a function of magnetic field for $\nu > 2.1$, indicating the presence of a well-defined Fermi surface). 

Figure~\ref{Fig2}a reveals the absence of distinct spectral features across this
photon energy range. This broadband character indicates that the photoresistance response is
governed by electronic heating, transduced through the strong temperature dependence
of $R_{\mathrm{xx}}$, rather than by a resonant process tied to a specific photon
energy. We emphasize, however, that the flatness of the spectral response constrains
the transduction mechanism but does not by itself identify the microscopic absorption
pathway: interband absorption followed by ultrafast carrier thermalization~\cite{mehew2024ultrafast} deposits
energy into the electronic subsystem and would produce an equivalent bolometric signal~\cite{Hubmann_2023}.
In either case, the absorbed power raises $T_\mathrm{e}$ and suppresses the correlated order.
Furthermore, to test the coupling mechanism, our device was endowed with a broadband
bow-tie antenna designed for linearly-polarized light. We found that the measured
$R_{\mathrm{THz}}$ is independent of the incident radiation polarization (see Supplementary Figures 5 and 6), suggesting that the response arises from direct absorption~\cite{Pol-not-pol} (see Supplementary Notes 3, 4 and 12).

As temperature increases (Fig.~\ref{Fig2}b–e), the photoresistance amplitude decreases gradually but remains observable up to $T_0 \approx 6$--$7~\mathrm{K}$, demonstrating that the bolometric effect persists well above the base temperature of 1.7~K. Furthermore, measurements in perpendicular magnetic fields reveal that the photoresponse remains robust up to several tesla, with substantial reduction occurring only above 4~T. 

We have also measured $R_\mathrm{THz}$ as a function of radiation power density, $\rho$. Upon raising the latter, the maximum $R_\mathrm{THz}$ at the CI state increases systematically, consistent with the elevated electronic temperature change $\Delta T_\mathrm{e}$ (shown in Fig.~\ref{Fig3}a). In principle, one can use DC heating that allows us to generate the same $\Delta T_\mathrm{e}$ in a controlled manner: by sweeping $I_\mathrm{DC}$, we gradually raise $T_\mathrm{e}$ and monitor the resulting suppression of the differential resistance $\frac{\mathrm{d}V}{\mathrm{d}I} (P_\mathrm{abs})$, where $P_\mathrm{abs}$ is the absorbed power.  
Figure~\ref{Fig3}a displays the filling-factor dependence of the $R_\mathrm{THz}$ alongside the quantity $\frac{\mathrm{d}V}{\mathrm{d}I} (P_\mathrm{abs}) - \frac{\mathrm{d}V}{\mathrm{d}I} (0)$ with each trace taken at fixed DC heating power $P_\mathrm{abs}$. This provides a direct measure of the absorbed power in the detector, a standard approach in bolometric techniques~\cite{han2013highly,kravtsov2025viscous}. The close correspondence between these two datasets demonstrates that both mechanisms -- THz absorption and DC heating - modify the resistance through the same underlying electronic temperature increase. 

\textbf{Performance.} 
Conventional bolometers are typically operated in the current or voltage bias regimes, so that instead of relating the incident power to the photoresistance, one measures $V_\mathrm{ph}=I_\mathrm{DC}\cdot R_\mathrm{THz}$. Thus, to evaluate the bolometric performance under operating conditions, we measure $V_\mathrm{ph}$ as a function of $I_\mathrm{DC}$ in proximity to the CI state (Fig.~\ref{Fig3}c) at the laser modulation frequency (see Methods). By such measurements, one extracts the difference between the $I$--$V$ curves measured in the dark and under THz excitation (Inset of Fig.~\ref{Fig3}e).  The strongest photovoltage response, as expected, emerges at $\nu=2$, where the CI exhibits the greatest temperature sensitivity, and the optimal operating point occurs at a bias current of $I^* = 80$ nA, as evident from the trace of $V_\mathrm{ph}(I_\mathrm{DC})$ at $\nu=2$ (Fig.~\ref{Fig3}d). 

Using the power calibration curve from Fig.~\ref{Fig3}b, we next plot $|V_\mathrm{ph}|$  as a function of absorbed power $P_\mathrm{abs}$ (Fig.~\ref{Fig3}e). In the linear regime, the voltage responsivity reaches $R^\mathrm{abs}_\mathrm{V} = (1.7\pm0.4)\cdot10^7$ $\mathrm{V/W}$, highlighting the extreme intrinsic sensitivity of the CI to the $T_\mathrm{e}$ increase (see also Supplementary Figure 3). For practical comparison, however, it is also instructive to estimate the external responsivity, defined with respect to the incident power. Since it is hard to estimate the absorption area, we use the responsivity per power density as the main practical figure of merit for the current device, which reaches a value of $2 \cdot 10^{-4}$~$\mathrm{V\cdot mm^{2}/\mu W}$, and potentially could be substantially improved through better absorption engineering (see the noise-equivalent power (NEP) estimate and full detector characterization in Supplementary Table 1). We further note that although electron heating should occur in any TBG device, such a remarkably high internal responsivity is expected to be observed only in high-quality stacks. This is because the fragile CI state must exhibit a pronounced insulating behavior, which varies significantly from device to device (see Supplementary Notes 6 and 7 for the response at filling factor $\nu=-2$ and measurements on additional MATBG devices, respectively). 

Finally, we consider the response time of the CI bolometers. In MATBG, carrier thermalization was experimentally shown to be governed by ultrafast electron--phonon Umklapp cooling, which establishes a picosecond intrinsic timescale for the photoresponse~\cite{mehew2024ultrafast}. In the Supplementary Information, we demonstrate non-linear dependence of electron overheating on the absorbed power, indicating electron--phonon cooling as the dominant heat escape mechanism. In our devices, however, the experimentally accessible speed is likely limited by the electrical RC time constant, $\tau_{\mathrm{RC}}$, due to the high device resistance. Because the structure is fabricated on an insulating substrate, the dominant capacitance originates from the graphene--graphite back-gate geometry. Using a typical operating resistance $R \sim 200~\mathrm{k}\Omega$ at the point $I^*$ and estimating the capacitance as $C \approx 0.25~\mathrm{pF}$ -- assuming an upper-bound device area of $250~\mu\mathrm{m}^2$ including the gated graphite leads - we estimate $\tau_{\mathrm{RC}} \sim50~\mathrm{ns}$ that in principle can be reduced by at least an order of magnitude in smaller devices. This value therefore sets the practical upper limit on the response speed in the present device geometry.

\section*{Discussion}
Our experiments can be further generalized for time-domain interrogation of the CI state itself. Radiation-driven excitation confined within the flat band, combined with time-resolved monitoring of the CI photoresistance, would allow direct access to the intrinsic timescale of correlated-order suppression and recovery, thereby probing the collective dynamics within the ground state. Unlike previous optical studies relying on high-energy interband excitation, such measurements would isolate the response of the many-body order parameter to low-energy perturbations and provide a new route to studying collective modes in moiré quantum materials. In addition to fundamental importance, the demonstrated internal responsivity exceeds that of many commercial bolometers in the same spectral window~\cite{sizov2018terahertz}. Depending on the wavelength, integration with cavities or metamaterials, plasmonic resonators, or photonic-crystal architectures, routinely employed in the field, could raise absorption toward unity and plausibly elevate responsivities beyond $10^{7}\,\mathrm{V/W}$ without compromising the nanosecond-scale response. Moreover, unlike superconducting detectors, such devices remain operational in magnetic fields up to several tesla, enabling compatibility with magneto-optical instrumentation. Looking forward, alternative correlated phases at different fillings, multilayer twisted stacks, their wafer-scale growth~\cite{Sun2021}, and integrated cavity geometries may support both single-pixel sensors and large-format arrays. 

\section*{Methods}
\textbf{Sample fabrication.} Our samples were fabricated from two graphene flakes cut using an AFM tip and subsequently
encapsulated between hBN slabs using a standard dry-transfer technique\cite{wang2013one}. The relative twist angle
between the two graphene layers was set during assembly with the aid of an optical microscope
equipped with micromanipulators and a high-precision rotation stage. An additional graphite layer
was attached to the TBG to serve as a graphite contact for the TBG.
Narrow graphite was attached to the bottom surface of the heterostructure to serve as a local
back gate. The completed stack was released onto an undoped, insulating Si/SiO$_2$ substrate to suppress reflections of incident THz radiation. Finally, standard electron-beam lithography, reactive
ion etching, and thin-film metal deposition were used to define the contact leads.

\textbf{Transport and photoresponse measurements.} The samples were measured in a Quantum Design OptiCool magneto-optical cryostat with a base temperature $T_0=1.7~$K. Terahertz radiation at 0.14~THz was generated by a Terasense source, while higher frequencies (2.5 and 3.5~THz) were produced using a quantum cascade laser (Lytid QCL). The radiation was guided to the device via an optical path consisting of two lenses, two mirrors, and a beam splitter, with the final lens mounted on a motorized XYZ stage to optimize focusing using the photoresponse signal as feedback (see Supplementary Figure 2). 

Low-noise transport measurements were performed using standard
lock-in amplifiers. All electrical contacts were equipped with cryogenic low-pass RC filters to suppress noise originating from the measurement electronics. An AC excitation current, $I(t)=I_\mathrm{AC}\sin(2\pi f_\mathrm{I} t)$, was supplied by a low-noise
voltage-controlled current source, where $I_\mathrm{AC}=10$~nA is the current amplitude and $f_\mathrm{I}=4$~Hz is the excitation frequency.  The DC bias (voltage and/or current) and the gate voltage were applied using a low-noise source-meter. 

Carrier concentration of our device was controlled by the electrostatic back gate, where $n$ is given as:
\begin{equation}
n_{} =
\frac{C_{\mathrm{BG}}}{e}
\left( V_{\mathrm{BG}} - V_{\mathrm{BG}}^{0} \right)
\end{equation}
where $C_{\mathrm{BG}}$ is the bottom-gate capacitance per unit area; $V_{\mathrm{BG}}^{0}$ denotes the corresponding charge-neutrality voltage, where $e$ is the elementary charge.

The photoresistance was defined as
\begin{equation}
 R_{\mathrm{THz}}(\nu) = R_{\mathrm{on}}(\nu) - R_{\mathrm{off}}(\nu) .
\end{equation}
Rather than measuring $R_{\mathrm{on}}(\nu)$ and $R_{\mathrm{off}}(\nu)$ 
separately, we employed a dual-modulation technique \cite{shilov2024high, kravtsov2025viscous} in which the THz radiation was 
square-wave modulated at $f_{\mathrm{mod}}=9$~Hz while simultaneously modulating the 
current at $f_\mathrm{I}$. The measured voltage can be written as
\begin{equation}
V(t) = I(t)R_{\mathrm{xx}}(t) + V_{\mathrm{ph}}(t),
\end{equation}
where $V_{\mathrm{ph}}(t)$ is the photovoltage oscillating at $f_{\mathrm{mod}}$.
Since the electronic temperature relaxation time is much shorter than 
$1/f_{\mathrm{mod}}=111$~ms, both $R_{\mathrm{xx}}$ and $V_{\mathrm{ph}}$ follow the 
modulation quasi-instantaneously.

The signal was analyzed using a lock-in amplifier operating in dual mode at the 
difference frequency $f_{\mathrm{dm}}=f_{\mathrm{mod}}-f_\mathrm{I}$. Fourier decomposition of the square-wave-modulated resistance yields
\begin{equation}
V_{\mathrm{dm}} = \frac{I_0\,\Delta R_{\mathrm{xx}}}{\pi},
\end{equation}
allowing direct extraction of the photoresistance,
\begin{equation}
\Delta R_{\mathrm{xx}} = \pi \frac{V_{\mathrm{dm}}}{I_0}.
\end{equation}

Photovoltage measurements were performed under DC bias using lock-in detection at 
$f_{\mathrm{mod}}$. The peak photovoltage was obtained from the RMS lock-in output 
$V_{\mathrm{ph}}^{\mathrm{lock\text{-}in}}$ via
\begin{equation}
V_{\mathrm{ph}} = \frac{\pi}{\sqrt2}\,V_{\mathrm{ph}}^{\mathrm{lock\text{-}in}},
\end{equation}
accounting for square-wave modulation and RMS-to-peak conversion.

\section*{Data Availability}
The data shown in Figs. 1–3 are provided in the Source Data files. All other data that support the findings of this study are available from the corresponding author upon request.

\newpage

\section*{Acknowledgements}
We thank Jeong Min Park and Pablo Jarillo-Herrero for the discussion on the correlated gap in MATBG. We acknowledge Ryo Mizuta Graphics as the source of the 3D optical-component graphics used in Supplementary Figure~2a.

\section*{Funding statement}
This work is supported by the National Research Foundation, Singapore under its NRF Fellowship (NRF Award NRF-NRFF17-2025-0007 given to D.A.B.). K.S.N. is grateful to the Ministry of Education, Singapore (Research Centre of Excellence award to the Institute for Functional Intelligent Materials, I-FIM, project number EDUNC-33-18-279-V12) and to the Royal Society (UK, grant number RSRP R 190000) for support. K.W. and T.T. acknowledge support from the JSPS KAKENHI (Grant Numbers 21H05233 and 23H02052), the CREST (JPMJCR24A5), JST and World Premier International Research Center Initiative (WPI), MEXT, Japan.

\section*{Author contributions}
D.A.B. conceived and supervised the project. L.E., A.L.S., M.K., X.Z., and S.J. performed photoresponse measurements. L.E., A.L.S. and S.J. contributed to the fabrication of devices. M.L., A.K. performed antenna absorption simulations. R.I., K.S., and I.G. performed noise thermometry measurements. L.E., A.L.S., M.K. analyzed the data with input from D.A.B.. L.E. and D.A.B. wrote the paper with input from all authors. G.N.G. and K.S.N. provided experimental support. A.P., and I.I. provided theory support. T.T. and K.W. provided high-quality hBN crystals. 
\section*{Competing Interests}
The authors declare no competing interests.
\section*{Figure Captions}

\makeatletter\close@column\makeatother
\clearpage
\makeatletter
\input{aps11pt4-1.rtx}
\@booleanfalse\twocolumn@sw
\@booleantrue\preprintsty@sw
\def\title@column#1{%
  \minipagefootnote@init
  #1%
  \minipagefootnote@foot
}
\def\close@column{\newpage}
\oddsidemargin=0pt
\evensidemargin=0pt
\topmargin=-37pt
\headheight=12pt
\headsep=25pt
\topskip=10pt
\footskip=30pt
\textheight=665.5pt
\textwidth=468pt
\columnwidth=468pt
\hsize=468pt
\linewidth=468pt
\columnsep=10pt
\def\baselinestretch{1.5}
\normalsize
\makeatother

\setcounter{page}{1}
\setcounter{section}{0}
\setcounter{subsection}{0}
\setcounter{figure}{0}
\setcounter{table}{0}
\setcounter{equation}{0}
\setcounter{suppnote}{0}
\setcounter{affil}{0}

\renewcommand*{\Im}{\operatorname{Im}}
\renewcommand*{\Re}{\operatorname{Re}}
\renewcommand{\thetable}{\arabic{table}}
\renewcommand{\tablename}{Supplementary Table}
\renewcommand{\theequation}{S\arabic{equation}}
\renewcommand{\thesubsection}{S\Roman{section}.\Roman{subsection}}
\renewcommand{\thefigure}{\arabic{figure}}
\renewcommand{\figurename}{Supplementary Figure}

\let\title\MergedTitleCommand
\let\author\MergedAuthorCommand
\let\affiliation\MergedAffiliationCommand

\newcommand{\SupplementaryMakeTitle}{%
  \let\MergedOriginalLabel\label
  \renewcommand{\label}[1]{\MergedOriginalLabel{suppfront:##1}}%
  \MergedMakeTitleCommand
  \thispagestyle{empty}%
  \let\label\MergedOriginalLabel
}
\let\maketitle\SupplementaryMakeTitle

\title{\LARGE{Supplementary Information for}\\\Large{Correlated Insulator Moiré Bolometer}}

\author{L. Elesin$^{+}$}
\affiliation{Department of Materials Science and Engineering, National University of Singapore, 117575, Singapore}

\author{A. L. Shilov$^{+}$}
\affiliation{Department of Materials Science and Engineering, National University of Singapore, 117575, Singapore}

\author{M. Kravtsov$^{+}$}
\affiliation{Department of Materials Science and Engineering, National University of Singapore, 117575, Singapore}

\author{X. Zhou}
\affiliation{Department of Materials Science and Engineering, National University of Singapore, 117575, Singapore}

\author{M. Lukianov}
\affiliation{Center for Neurophysics and Neuromorphic Technologies, Moscow, 127495, Russia}

\author{A. Kuksov}
\affiliation{Center for Neurophysics and Neuromorphic Technologies, Moscow, 127495, Russia}

\author{S. Jana}

\affiliation{Department of Materials Science and Engineering, National University of Singapore, 117575, Singapore}

\author{I. Iorsh}
\affiliation{Queen's University, Kingston, K7L 3N6, Canada}

\author{R. Izmaylov}
\affiliation{Moscow Pedagogical State University, Moscow, 119991, Russia}

\author{K. Shein}
\affiliation{Moscow Pedagogical State University, Moscow, 119991, Russia}

\author{I. Gayduchenko}
\affiliation{Moscow Pedagogical State University, Moscow, 119991, Russia}

\author{T. Taniguchi}
\affiliation{International Center for Materials Nanoarchitectonics, National Institute for Materials Science, Tsukuba, 305-0044, Japan}

\author{K. Watanabe}
\affiliation{Research Center for Functional Materials, National Institute for Materials Science, Tsukuba, 305-0044, Japan}

\author{K. S. Novoselov}
\affiliation{Institute for Functional Intelligent
Materials, National University of Singapore, Singapore, 117575, Singapore}

\author{G. N. Goltsman}
\affiliation{Moscow Pedagogical State University, Moscow, 119991, Russia}

\author{A. Principi}
\affiliation{School of Physics and Astronomy, University of Manchester, Manchester, M13 9PL, United Kingdom}

\author{D. A. Bandurin$^{*}$}
\affiliation{Department of Materials Science and Engineering, National University of Singapore, 117575, Singapore}
\affiliation{Institute for Functional Intelligent
Materials, National University of Singapore, Singapore, 117575, Singapore}

\maketitle

\newpage
\begin{center}
    \textbf{CONTENTS}
\end{center}

\vspace{1em}

\begin{enumerate}
    \setlength{\itemsep}{0.7em}

    \item Extended transport data

    \item Incident power calibration and responsivity estimates

    \item Polarization dependence of a reference device

    \item Polarization independence of photoresistance in MATBG sample

    \item Comparison of photoresistance with DC heating

    \item Photoresistance at $\nu=-2$

    \item Measurements of additional MATBG samples

    \item NEP, DR, and $G_{\mathrm{th}}$ estimation

    \item THz-induced electron heating and heat transfer balance

    \item Noise thermometry in TBG

    \item Lattice temperature

    \item Electromagnetic simulations

    \item Correlated insulator vs superconducting bolometry

    \item CI phase diagram

\end{enumerate}

\newpage

\suppnote{Extended Transport Data}

\begin{figure*}[ht!]
  \centering\includegraphics[width=1\linewidth]{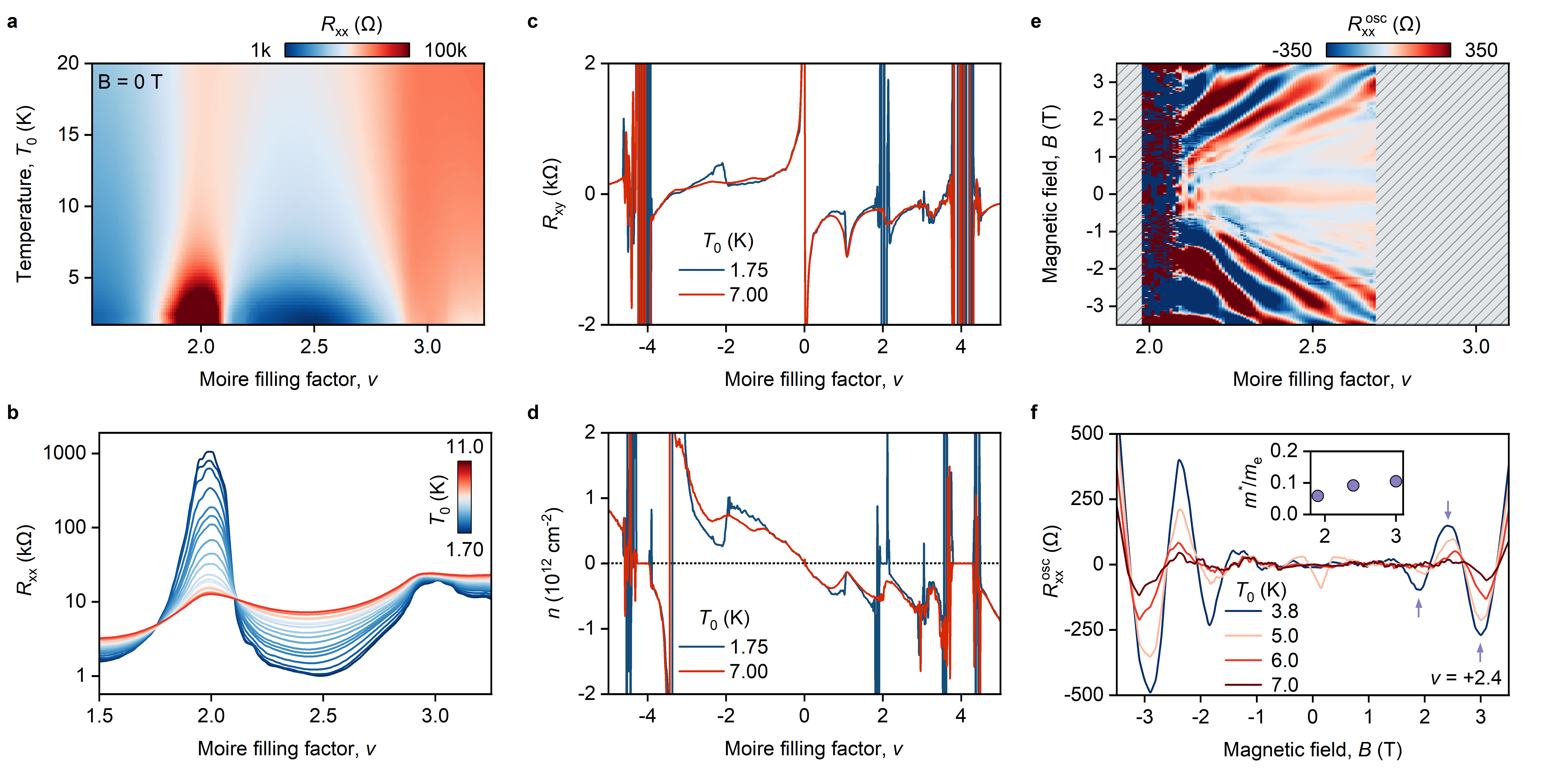}
    \caption{\textbf{Transport data in a voltage-bias scheme}. 
\textbf{a}, Longitudinal resistance map as a function of temperature and moiré filling factor. The measurement scheme employs a small AC bias voltage $U$, which prevents additional electron heating, since in the correlated state the dissipated power scales as $U^{2}/R$ rather than $I^{2}R$, as in a current-bias scheme.
\textbf{b}, Selected traces of the dependence of $R_\mathrm{xx}$ on $\nu$. 
\textbf{c}, Hall resistance as a function of $\nu$ antisymmetrized with respect to $B = \pm 200$~mT for selected temperatures $T_\mathrm{0}$. 
\textbf{d}, Carrier concentration extracted from the Hall resistance ($n=\frac{B}{R_\mathrm{xy}e}$) at selected temperatures. 
\textbf{e}, Extracted oscillatory component of the longitudinal resistance as a function of magnetic field and filling factor $\nu$. 
\textbf{f}, Line cuts of the oscillatory longitudinal resistance map at different temperatures. Inset: effective carrier mass extracted from Lifshitz–Kosevich (LK) fits ($m_\mathrm{e}$ -- free electron mass).}

	\label{transport}
\end{figure*}

\newpage

\suppnote{Incident power calibration and responsivity estimates} 
To determine the external responsivity of the device, the optical power incident on the device area was calibrated by accounting for losses in the optical path using a pyroelectric sensor (PS) placed in the same optical configuration as that used for the MATBG measurements, namely at the focal plane of Lens 2.

For measurements at 0.14 THz, the PS was first calibrated against the full output power of the THz source, denoted $P_0$. With the PS aperture of diameter $D_\mathrm{s}$ fully illuminated (the PS positioned directly in front of the 0.14 THz waveguide), the corresponding detector signal $V_0$ was taken as a reference. The complete optical setup (see  Supplementary Figure~\ref{setup}a) was then reconstructed, and the PS was placed at the same position as the MATBG device during the experiments and scanned laterally across the beam.

At each position $(x,y)$, the measured PS signal $\Delta V(x,y)$ is proportional to the optical power incident on the aperture at that location $P(x,y)$:
\begin{equation}
P(x,y) = P_0 \frac{\Delta V(x,y)}{V_0}.
\end{equation}
Dividing $P(x,y)$ by the aperture area $S_\mathrm{PS} $ yields the averaged local power density $\rho(x,y) = P(x,y)/S_\mathrm{PS}$ in the device plane, which we assume to be the same within the aperture size. By repeating this procedure over the scan area with a step size $\delta x = \delta y = D_\mathrm{s}/2$, we reconstruct the spatial beam profile and extract the effective beam area (see Supplementary Figure~\ref{setup}c). The resulting beam exhibits a characteristic diameter $D$ much larger than $D_\mathrm{s}$: $D \gg D_\mathrm{s}$, which justifies the approximation of a uniform power density across the PS aperture. The maximum value of $\rho (x,y)$ was taken as the value of the maximum incident power density $\rho_0=70 ~\mu\mathrm{W/mm}^2$ incident on the MATBG during the experiments (see Supplementary Table~\ref{tab:power_responsivity}).

At 3.5 THz, the radiation source (QCL) emits a strongly divergent beam at the output ($\sim30\degree$), which prevents a direct reproduction of the reference measurement $V_0$ before the collimating lens used for 0.14 THz. Therefore, an alternative method for estimating the power density was employed. Since the pyroelectric sensor was factory-calibrated at these high frequencies, we used the manufacturer-provided calibration curve $P = P(\Delta V)$ to convert the measured sensor signal $\Delta V(x,y)$ in the device plane into the optical power incident on the PS, $P(x,y)$.

Using a THz camera (see inset of Supplementary Figure~\ref{setup}b), we determined that the focused 3.5 THz beam has a characteristic diameter $D = 1.5~\mathrm{mm} \ll D_\mathrm{s}$. Because the beam size is smaller than the PS aperture during the lateral $(x,y)$ scan, the sensor collects the entire beam power. To accurately estimate the maximum power density $\rho_0$, we used a small scanning step along $(x, y)$: $\delta x = \delta y \ll D \ll  D_\mathrm{s}$. When the beam from 3.5 THz falls within the aperture area, we have practically the same value of $ P(x,y) = \mathrm{const}$, and when the beam goes beyond the aperture, $P(x,y)$ becomes zero. Therefore, the characteristic size of the recorded signal in this case is set by the PS aperture, with an effective diameter $D_\mathrm{s} = 6\,\mathrm{mm}$ (see Supplementary Figure~\ref{setup}b). The scan position at which the measured power $P(x,y)$ reaches its maximum was used to estimate the maximum power density,
\begin{equation}
\rho_0 = \frac{\max\!\left[P(x,y)\right]}{S},
\end{equation}
where $S$ is the area of the focused 3.5~THz beam, determined from the beam photograph.

Knowing the maximum power density $\rho_{0}$ for each radiation frequency used in the experiment, we estimate the optical responsivity, defined as the ratio of the measured photovoltage to the incident optical power in the linear response regime. 
First, we define the voltage responsivity per incident power density,

\begin{equation}
    R_\mathrm{V}^{\rho} \equiv \frac{V_{\mathrm{ph}}}{\rho_{\mathrm{}}},
\end{equation}
where $\rho_{\mathrm{}}$ is the incident power density at the sample 
position. This quantity is directly measurable and requires no assumptions about
focusing quality or effective collection area, making it the natural,
assumption-free figure of merit for a device small compared to the
wavelength, for which the antenna is inactive. Using the 
power-density calibration, we 
obtain in the linear regime

\begin{equation}
    R_\mathrm{V}^{\rho}(\lambda = 85~\mu\mathrm{m})
    \sim
    3.2 \times 10^{-6}
    ~\frac{\mathrm{V\,m^2}}{\mathrm{W}},
\end{equation}

\begin{equation}
    R_\mathrm{V}^{\rho}(\lambda = 2140~\mu\mathrm{m})
    \sim
    2.1 \times 10^{-4}
    ~\frac{\mathrm{V\,m^2}}{\mathrm{W}}.
\end{equation}

Next, we convert $R_\mathrm{V}^{\rho}$ into a conventional voltage responsivity $R_\mathrm{V}$ --- photovoltage divided by total incident power -- for the diffraction-limited configuration. In our present setup, the beam 
diameter at the sample plane is approximately $2.5$~cm, which we verify 
against the diffraction limit. For $\lambda = 2140~\mu\mathrm{m}$, the 
effective numerical aperture is set by the ratio of the illuminated beam 
diameter on the focusing lens to twice the focal length,

\begin{equation}
    \mathrm{NA}
    \simeq
    \frac{D_{\mathrm{beam}}}{2f}
    =
    \frac{5~\mathrm{cm}}{2 \times 25~\mathrm{cm}}
    = 0.1,
\end{equation}

giving a diffraction-limited Airy diameter of

\begin{equation}
    d_{\mathrm{Airy}}
    =
    \frac{2 \times 0.61\,\lambda}{\mathrm{NA}}
    \simeq
    \frac{1.22 \times 2140~\mu\mathrm{m}}{0.1}
    \simeq
    2.6~\mathrm{cm},
\end{equation}
in good agreement with the measured beam size.  Although our setup operates at
the diffraction limit, the achieved focusing is far from ideal, as a direct
consequence of the limited $\mathrm{NA} = 0.1$ of the available optics.
Integrating the beam irradiance profile $\rho(x,y)$ over the device plane yields a total incident power of $P_{\mathrm{inc}}\simeq17$~mW, indicating that approximately $50\%$ of the nominal $30$~mW source power is lost along the optical path.

Finally, we estimate the achievable practical
responsivity for a well-focused setup (for $\lambda=2140~ \mu$m, calculations for $\lambda=85~ \mu$m are provided in the Supplementary Table~\ref{tab:power_responsivity}). By increasing the numerical aperture close
to $\mathrm{NA} \approx 1$, the focused beam diameter can be reduced by
approximately one order of magnitude relative to the present setup. Since the focal spot area scales as $d^{2}$, this corresponds to a reduction
in spot area -- and thus in the total incident power required to maintain
the same local power density -- of approximately two orders of magnitude.
We note that, even in the case of ideal focusing, the beam area remains much
larger than the total device area, so that the device's responsivity to
incident power density remains the relevant figure of merit. Since $R_\mathrm{V}^{\rho}$ is an intrinsic property of the
device, independent of focusing conditions, the same photovoltage would be
produced with only $P_{\mathrm{inc}}^{\,\mathrm{NA}=1} \sim 0.17$~mW,
yielding

\begin{equation}
    R_\mathrm{V}^{\,\mathrm{NA}=1}
    \sim
    10^2~\mathrm{V/W}.
\end{equation}

\begin{figure*}[ht!]
  \centering\includegraphics[width=0.65\linewidth]{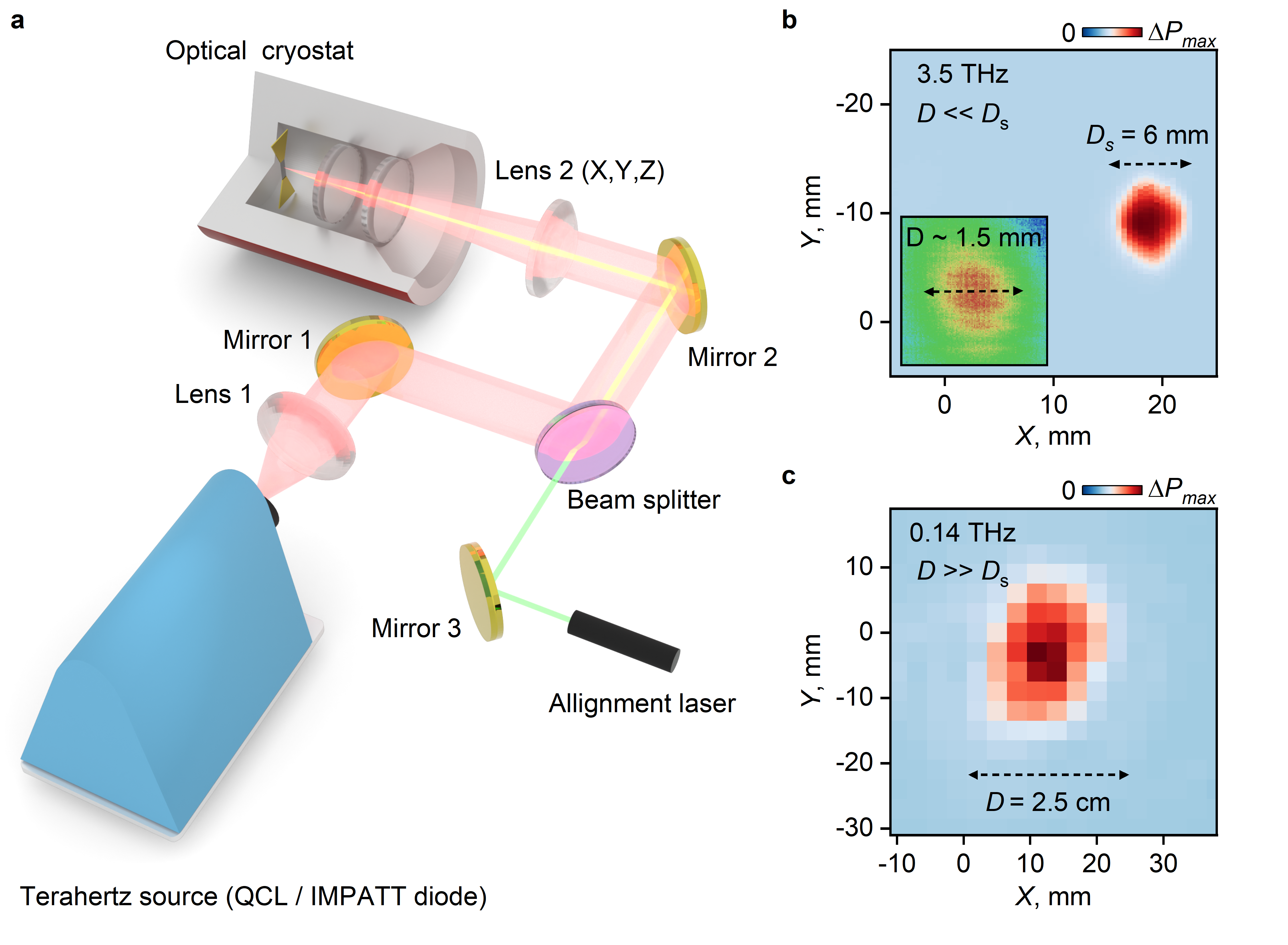}
    \caption{
    \textbf{Experimental setup and power calibration.} 
\textbf{a}, Optical path consisting of several optical elements, including a collimating lens (Lens 1), mirrors (Mirrors 1 and 2), a beam splitter, a focusing lens (Lens 2), and an alignment laser. The THz radiation passes through two optical cryostat windows (TPX and Tsurupica), which are transparent in the THz range. The 3D models of the optical elements were created by Ryo Mizuta Graphics.
\textbf{b}, Power calibration map of the 3.5 THz beam. The map represents the signal recorded by a commercial terahertz detector, which is proportional to the incident power. Inset: photograph of the beam captured using a 3.5 THz camera. 
\textbf{c}, Map of the power incident on the detector as a function of detector position for a 0.14 THz beam.
}
	\label{setup}
\end{figure*}

\begin{figure*}[ht!]
  \centering\includegraphics[width=\linewidth]{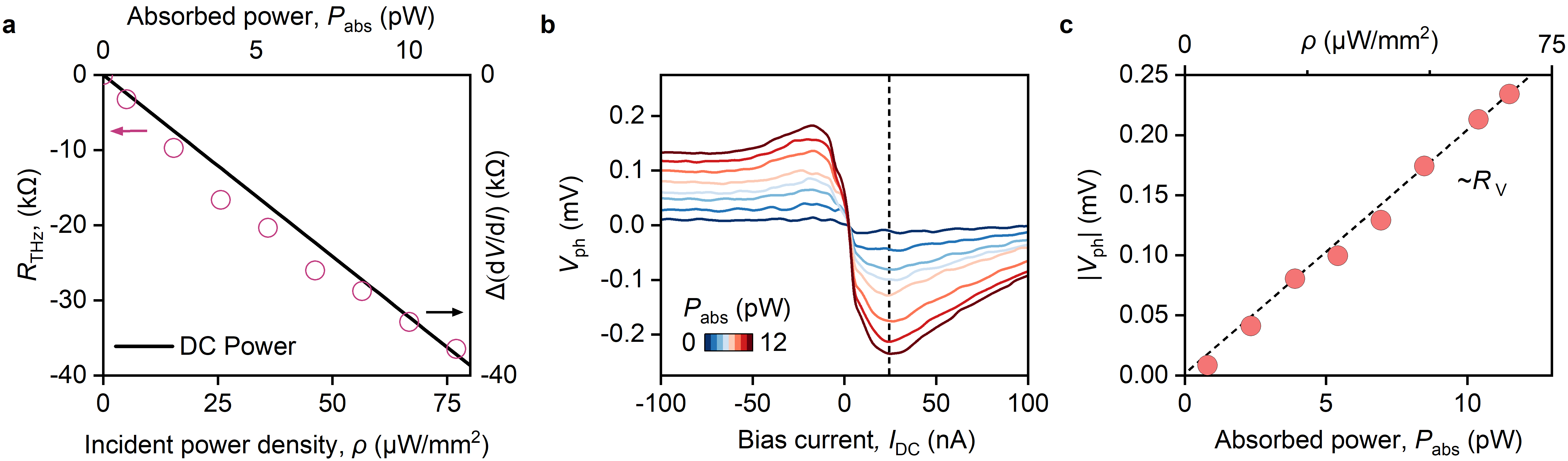}
\caption{\textbf{Bolometric response for 3.5 THz.} 
\textbf{a}, Comparison between the photoresistance at $\nu = 2$ as a function of the incident THz power density and the change in differential resistance, $\frac{\mathrm{d}V}{\mathrm{d}I}(P_{\mathrm{abs}}) - \frac{\mathrm{d}V}{\mathrm{d}I}(0)$, induced by DC heating with absorbed power $P_{\mathrm{abs}}$. 
\textbf{b}, $V_{\mathrm{ph}}$ as a function of the DC bias current $I_{\mathrm{DC}}$ under THz illumination at different power levels (the incident power is converted to absorbed power using the calibration curve). 
\textbf{c}, $|V_{\mathrm{ph}}|$ as a function of $P_{\mathrm{abs}}$ under THz illumination. The data were acquired at a fixed DC bias current of $I_{\mathrm{DC}} = 25\,\mathrm{nA}$. The black dashed line represents the device responsivity $R_{\mathrm{V}}$.}
	\label{3_5}
\end{figure*}

\newpage
\clearpage

\suppnote{Polarization dependence of a reference device}

In the polarization configuration aligned with the antenna axis, the incident terahertz electric field $E_0$ is expected to induce a potential difference between the antenna wings of the order $V \sim E_0 L$,
where $L$ is the effective length of the antenna. If this potential were efficiently coupled to the active device region, it would be dropped across the narrow channel of length $L_{\mathrm{dev}} \ll L$, leading to an enhanced local electric field $E_{\mathrm{loc}} \sim E_0 \frac{L}{L_{\mathrm{dev}}}$.
Since the absorbed power scales with the square of the electric field, this mechanism would result in an enhancement of the absorbed power by a factor of $\left(\frac{L}{L_{\mathrm{dev}}}\right)^2$
compared to a device without an antenna, as well as compared to the response under radiation with orthogonal polarization. 

\begin{figure}[ht!]
\centering
\begin{minipage}[c]{0.56\linewidth}
  \centering
  \includegraphics[width=0.85\linewidth]{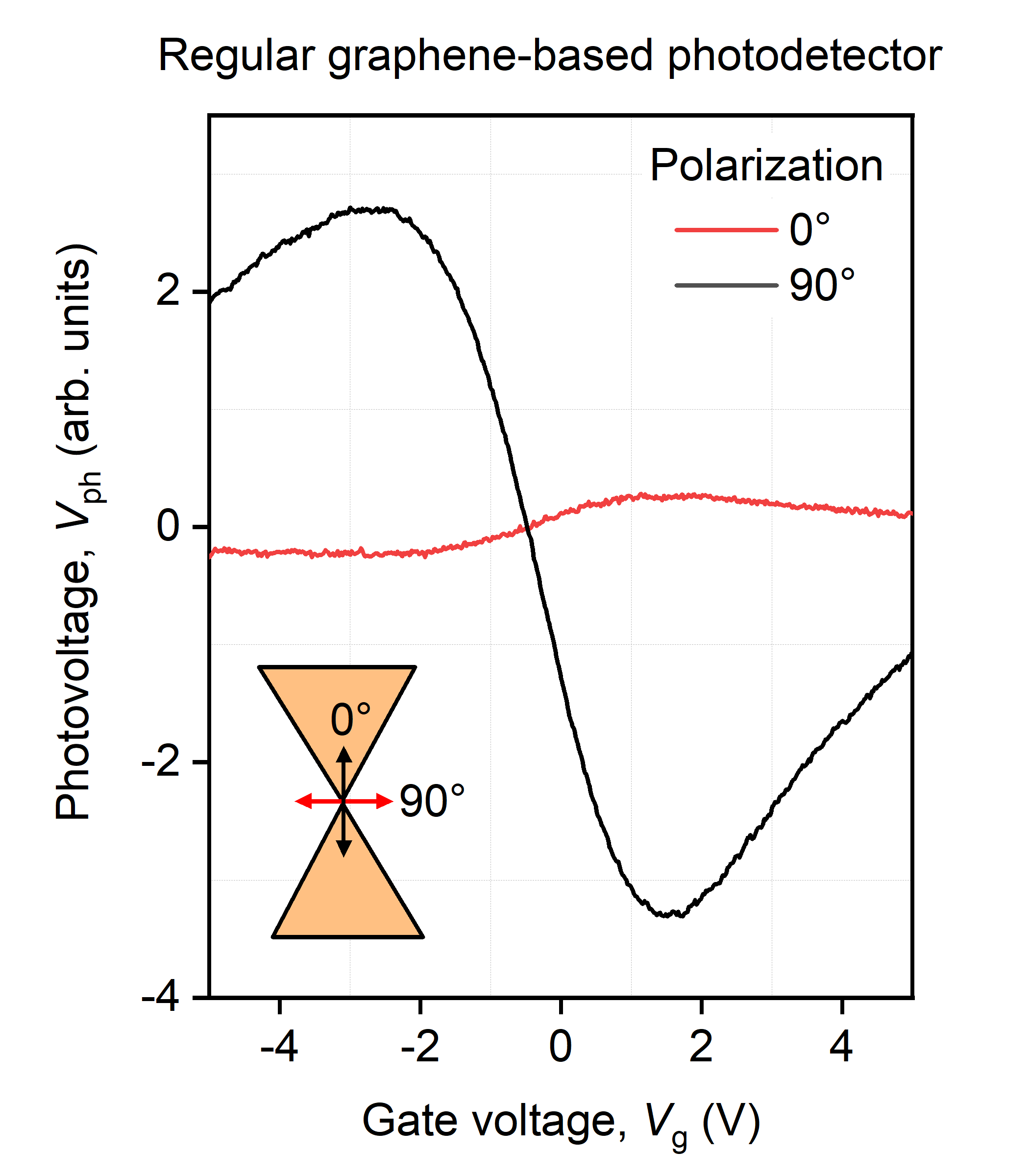}
\end{minipage}\hfill
\begin{minipage}[c]{0.40\linewidth}
  \caption{\textbf{Polarization dependence of a regular graphene-based photodetector equipped with a bow-tie antenna.} The photoresponse is maximized when the incident electric field is aligned with the antenna axis ($0\degree$) and is strongly suppressed for orthogonal polarization ($90\degree$).}
  \label{graphene_pn}
\end{minipage}
\end{figure}

Supplementary Figure~\ref{graphene_pn} shows a photovoltage of a reference monolayer graphene-based device endowed with the same bow-tie antenna as we used for our MATBG device for two orthogonal polarizations. Unlike the case of MATBG, a regular graphene photodetector exhibits expected suppression of the photovoltage when the radiation polarization is perpendicular to the antenna lobes. The absence of such a strong polarization-dependent enhancement in our MATBG bolometer indicates that antenna-mediated field enhancement does not contribute to the electron heating because of enormous impedance mismatch. 

\newpage

\suppnote{Polarization independence of photoresistance in MATBG sample}

\begin{figure}[ht!]
    \centering
    \includegraphics[width=1\linewidth]{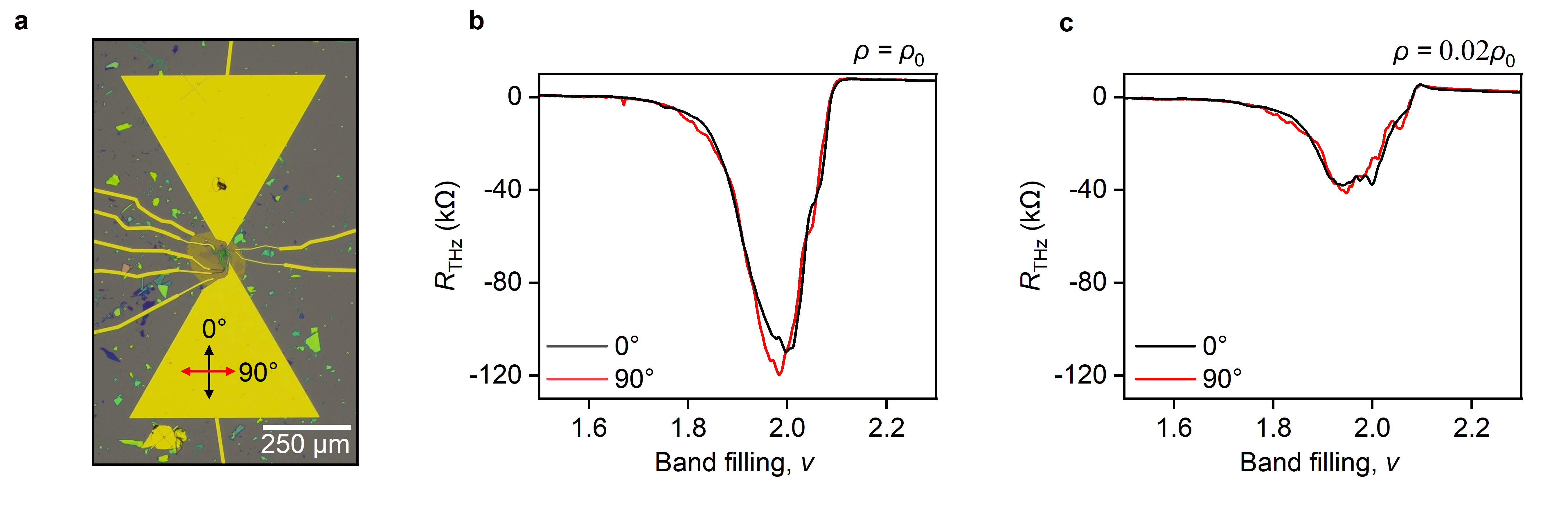}
    \caption{
    \textbf{Independence of the photoresistance on THz polarization.} 
    \textbf{a}, Optical microphotograph of the device. Arrows indicate the polarization of the incident terahertz radiation. 
\textbf{b-c}, $R_\mathrm{THz}$ measured with the incident polarization aligned with the bow-tie antenna and perpendicular to it, at the maximum output power density $\rho=\rho_0$ (\textbf{b}) and at an attenuated power density $\rho=0.02\rho_0$ in the linear regime (\textbf{c}). 
    }
    \label{Polar}
\end{figure}

For our MATBG device, we directly measured and compared the photoresistance at different polarization angles. The photoresistance traces obtained for two orthogonal polarizations, $R_{\mathrm{THz}}^{0}$ and $R_{\mathrm{THz}}^{90}$, nearly overlap across the entire range of band fillings (Supplementary Figure~\ref{Polar}). This demonstrates that the dominant photoresistive response cannot be attributed to polarization-selective antenna coupling and is instead consistent with a polarization-insensitive bolometric response arising from direct absorption in the MATBG channel.

Additionally, we compared the polarization-dependent change of the photoresistance with the total photoresistance signal. At both $0.14~\mathrm{THz}$ and $3.5~\mathrm{THz}$, the difference between photoresistance measured for the two orthogonal polarizations $\Delta R_\mathrm{THz}=R^0_\mathrm{THz}-R^{90}_\mathrm{THz}$ is orders of magnitude smaller than the photoresistance $R^0_\mathrm{THz}(R^{90}_\mathrm{THz})$ (see Supplementary Figure~\ref{Polar2}). 

\begin{figure}[ht!]
    \centering
    \includegraphics[width=1\linewidth]{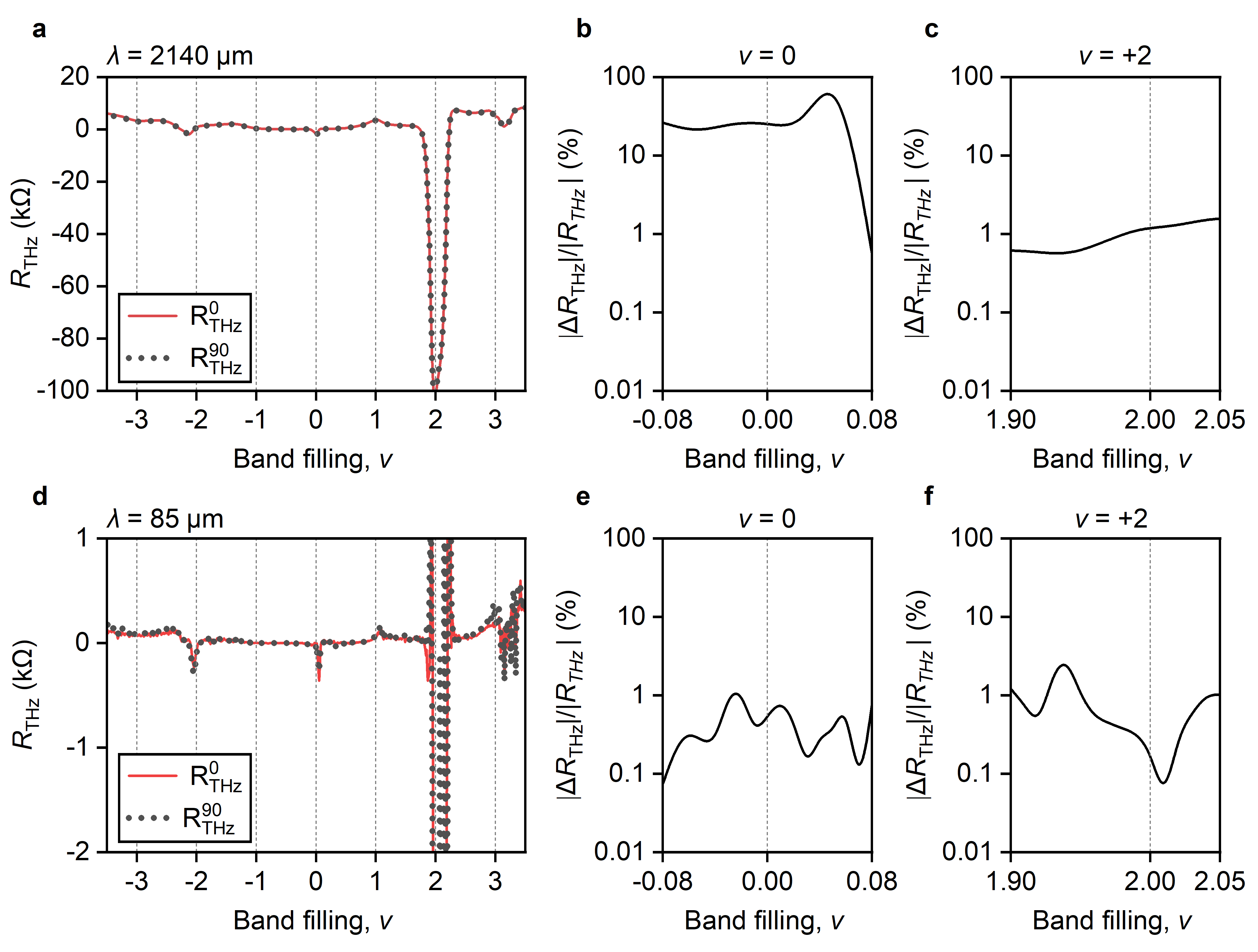}
    \caption{
    \textbf{Analysis of photoresistance on THz polarization.} 
    Comparison of the difference in the signal under 90 and 0 degree polarization for $\lambda=2140~\mu$m (\textbf{a-c}) and for $\lambda=85~\mu$m (\textbf{d-f}). 
    }
    \label{Polar2}
\end{figure}

\newpage
\clearpage
\suppnote{Comparison of photoresistance with DC heating}

Independently of photoresistance measurements, we measured the DC-bias-induced change in differential resistance in a 4-probe configuration for the same filling factors.
\begin{equation}
\Delta \left( \frac{\mathrm{d}V}{\mathrm{d}I} \right)(I_{\mathrm{DC}},\nu) = \frac{\mathrm{d}V}{\mathrm{d}I}(I_{\mathrm{DC}},\nu)-\frac{\mathrm{d}V}{\mathrm{d}I}(0,\nu)
\end{equation}
A comparison between $R_\mathrm{THz}$ and $\Delta (\frac{\mathrm{d}V}{\mathrm{d}I})$ at fixed \(I_{\mathrm{DC}}\) is not sufficient, because the dissipated DC power changes with filling factor through the two-terminal resistance. Therefore, we measured the full two-terminal differential-resistance map, $\frac{\mathrm{d}V}{\mathrm{d}I}_\mathrm{2pt}(I_{\mathrm{DC}},\nu)$
and reconstructed the two-terminal voltage as
\begin{equation}
V_\mathrm{2pt}(I_{\mathrm{DC}},\nu)
=
\int_0^{I_{\mathrm{DC}}}
\frac{\mathrm{d}V_\mathrm{2pt}}{\mathrm{d}I}(I',\nu)\,\mathrm{d}I' .
\end{equation}
The dissipated DC power was then calculated as
\begin{equation}
P_{\mathrm{abs}}(I_{\mathrm{DC}},\nu)
=
I_{\mathrm{DC}}V_\mathrm{2pt}(I_{\mathrm{DC}},\nu).
\end{equation}

\begin{figure}[ht!]
    \centering
    \includegraphics[width=1\linewidth]{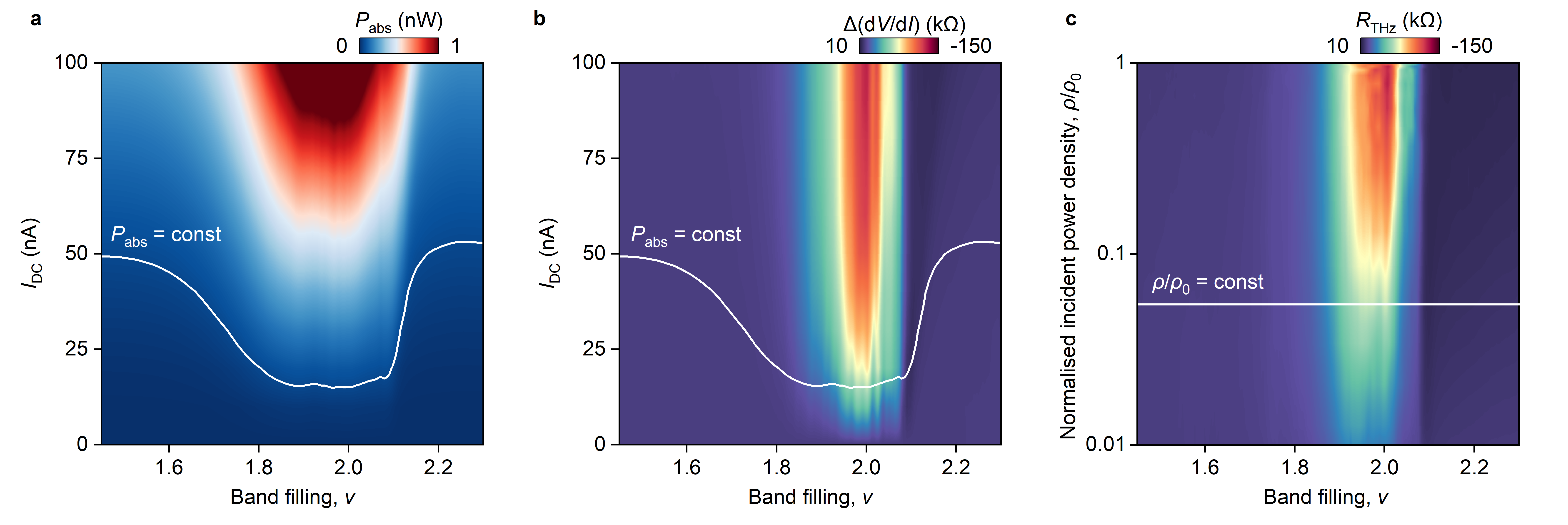}
    \caption{
    \textbf{Comparison between DC-heating and THz-induced response.}
\textbf{a}, Absorbed DC power, $P_\mathrm{abs}$, calculated from the measured two-terminal differential resistance as a function of band filling, $\nu$, and DC bias current, $I_\mathrm{DC}$. The white line marks an isopower contour, $P_\mathrm{abs}=\mathrm{const}$.
\textbf{b}, Change in differential resistance, $\Delta \left( \frac{\mathrm{d}V}{\mathrm{d}I} \right)$, measured under DC bias in the same parameter space. The same isopower contour is overlaid to extract the differential resistance change at fixed absorbed power.
\textbf{c}, THz-induced photoresistance, $R_\mathrm{THz}$, as a function of band filling and normalized incident power density, $\rho/\rho_0$. The white horizontal line indicates a constant incident power density, $\rho=\mathrm{const}$, used for comparison with the DC-heating response.}
    \label{constant_power}
\end{figure}

This gives a two-dimensional map \(P_{\mathrm{abs}}(I_{\mathrm{DC}},\nu)\) shown in Supplementary Figure~\ref{constant_power}a. From this map, we numerically extracted contours of constant power, \(P_{\mathrm{abs}}=\mathrm{const}\), and used these isopower traces to plot
\begin{equation}
\Delta \left( \frac{\mathrm{d}V}{\mathrm{d}I} \right)(\nu;P_\mathrm{abs})
=
\frac{\mathrm{d}V}{\mathrm{d}I}(I_{\mathrm{DC}}(\nu;P_\mathrm{abs}),\nu)-\frac{\mathrm{d}V}{\mathrm{d}I}(0,\nu).
\end{equation}
using the map $\Delta \left( \frac{\mathrm{d}V}{\mathrm{d}I}\right)(I_\mathrm{DC};\nu)$ shown in the Supplementary Figure~\ref{constant_power}b.

Next, we compared these DC-heating traces with the THz photoresistance traces captured at constant incident power densities (see Supplementary Figure~\ref{constant_power}c). The resulting comparison of traces is shown in Figure 3a in the main text. The conversion between incident THz power and effective absorbed power was determined by matching the $R_\mathrm{THz}(\rho)$ with $\Delta \left(\frac{\mathrm{d}V}{\mathrm{d}I}\right)(P_\mathrm{abs})$ for $\nu = 2$. This calibration was performed at the base temperature and $B=0$, and provides an independent conversion between the incident THz power density and the effective absorbed power responsible for electronic overheating.

We then used the same conversion factor to compare the photoresistance measured as a function of incident THz power with the Joule-heating-induced change in $(\frac{\mathrm{d}V}{\mathrm{d}I})$ measured as a function of $P_{\mathrm{abs}}$ at different magnetic fields and temperatures. As shown in Supplementary Figure~\ref{calib}, the evolution of $\Delta(\frac{\mathrm{d}V}{\mathrm{d}I})$ under DC Joule heating closely reproduces the corresponding THz photoresistance over the measured range of $B$ and $T$. In other words, the calibration obtained at the base temperature and zero magnetic field remains valid when these external parameters are varied.

\begin{figure}[ht!]
    \centering
    \includegraphics[width=0.8\linewidth]{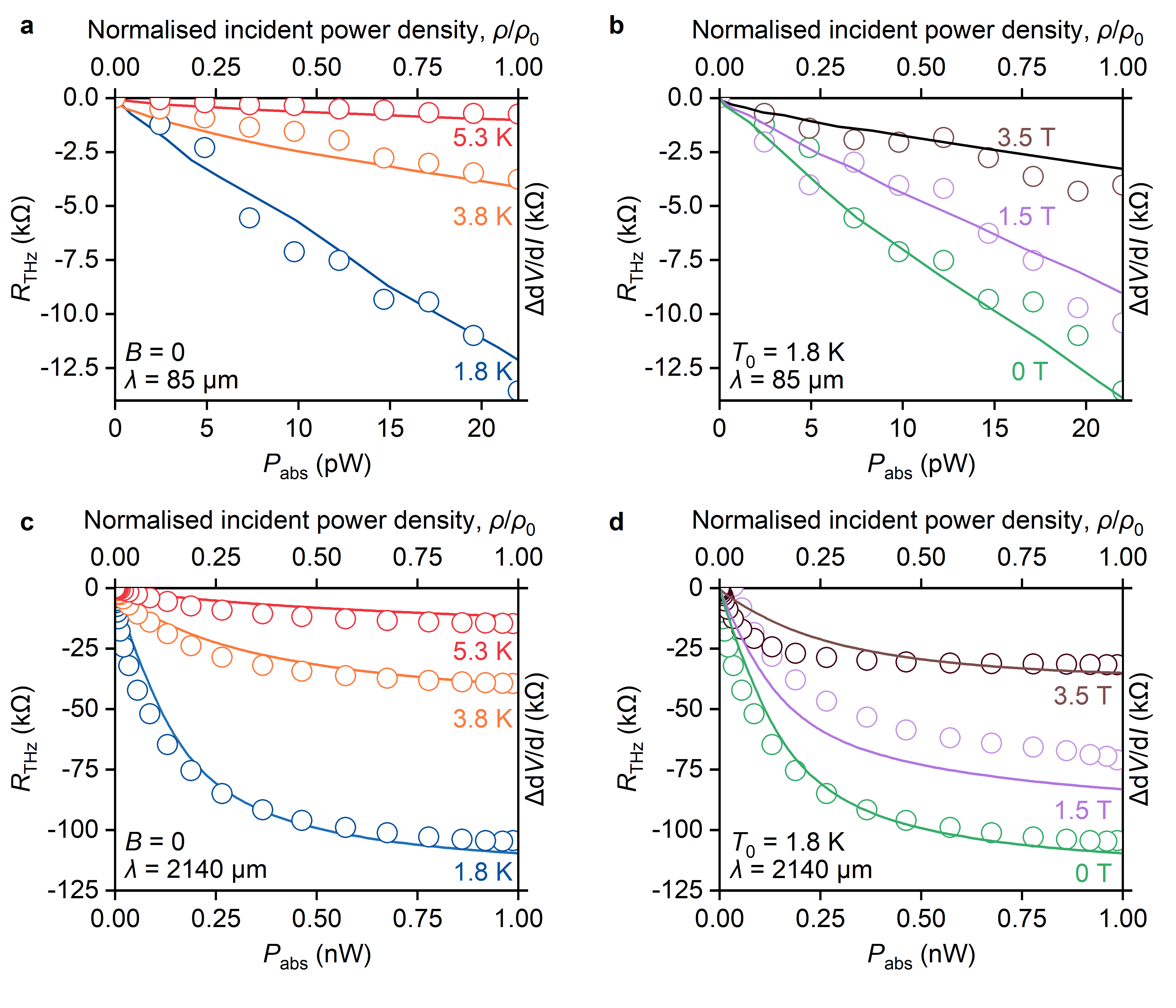}
    \caption {
    \textbf {Joule heating--radiation absorption analysis in $B$--$T_0$ phase space.} \textbf{a-b}, Comparison between the photoresistance at $\nu = 2$ as a function of incident power density of $\lambda=85~\mu$m (3.5 THz) source and the change in $\frac{\mathrm{d}V}{\mathrm{d}I}(P_{\mathrm{abs}})-\frac{\mathrm{d}V}{\mathrm{d}I}(0)$ produced by DC heating with power $P_{\mathrm{abs}}$ at different temperatures (\textbf{a}) and magnetic fields (\textbf{b}). Power absorbed by $\lambda=85~\mu$m source does not drive the response out of the linear regime. \textbf{c-d}, Same as \textbf{a-b} but for $\lambda=2140 ~\mu$m (0.14 THz) source. Solid lines represent $\Delta(\frac{\mathrm{d}V}{\mathrm{d}I})(P_{\mathrm{abs}})$, while circles show $R_{\mathrm{THz}}(\rho/\rho_0)$. The corresponding error bars are smaller than the symbol size and therefore lie within the circles.}
    \label{calib}
\end{figure}

\newpage
\clearpage

\suppnote{Photoresistance at $\nu=-2$}

In our interpretation, radiation-induced electronic heating can occur at all filling factors; however, a large photoresistance signal is expected only where the resistance has a strong temperature dependence, i.e. where $\mathrm{d}R\mathrm{/d}T$ is large. Therefore, the response is not specific to $\nu=+2$ in principle, but its magnitude is strongly enhanced at the most pronounced correlated-insulating states.

Consistent with this expectation, we also observe a negative photoresistance response near $\nu=-2$, where the transport data show insulating behavior. However, because the correlated-insulating feature at $\nu=-2$ is substantially weaker than that at $\nu=+2$, the corresponding photoresistance signal has the same sign and qualitative behavior but a smaller amplitude.

\begin{figure}[ht!]
    \centering
    \includegraphics[width=1\linewidth]{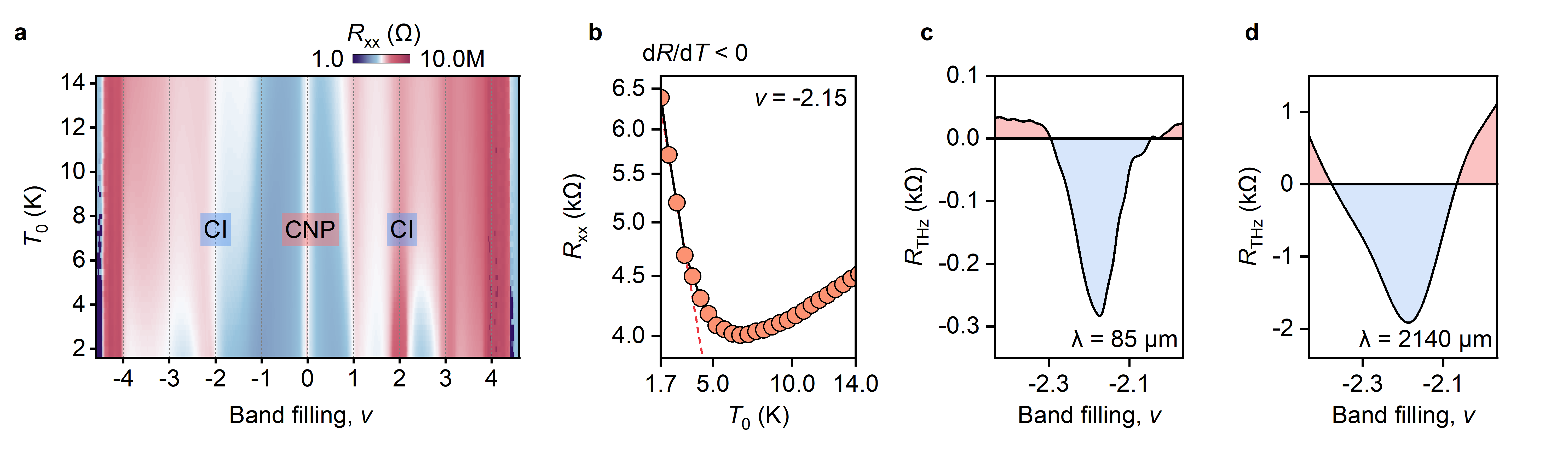}
    \caption{
    \textbf{Photoresistance analysis at the filling factor $\nu = -2$.}
\textbf{a}, Temperature dependence of the longitudinal resistance $R_{\mathrm{xx}}$ across moiré band fillings.
\textbf{b}, Insulating behavior extracted from transport measurements at $\nu=-2$.
\textbf{c-d}, Photoresistance near $\nu=-2$ under FIR and MM-wave irradiation.}
    \label{tbg_2}
\end{figure}

\newpage
\suppnote{Measurements of additional MATBG samples}
To assess the reproducibility of the bolometric photoresponse in MATBG we fabricated and measured two additional MATBG devices under MM/FIR irradiation. In all cases where the photoresponse could be reliably assessed, irradiation produced electronic heating, consistent with the hot-electron mechanism. 

It is important to emphasize that the giant internal responsivity is not expected in an arbitrary TBG device. It requires a well-developed correlated insulating state, because the response is governed by the temperature sensitivity of the resistance. Devices with weaker correlated features show a much smaller photoresistance and therefore a smaller photovoltage, even though electronic heating is still present. This is illustrated in Supplementary Figure~\ref{tbg}, where other MATBG devices ($\theta = 1.05^\circ$ and $\theta=1.09^\circ$) exhibit a negative photoresistance at the insulating state near $\nu = 2$ (for MATBG2) and $\nu=3$ (for MATBG3)~\cite{stepanov2020untying}. In both cases the amplitude is much smaller than in MATBG1 ($\theta = 1.01^{\circ}$), the device described in the main text. We also note that, for the candidate intervalley-coherent state (at $\nu=2$), a nonzero order parameter does not necessarily imply a large spectral gap: theoretical work has shown that disorder or strain can strongly suppress, or even eliminate, the spectral gap while intervalley coherence remains finite. Therefore, the relevant requirement for a giant bolometric response is not only the presence of correlated order, but also a sufficiently robust transport gap and a strong temperature dependence of $R_\mathrm{xx}$, and it naturally changes from device to device.

\begin{figure}[ht!]
    \centering
    \includegraphics[width=0.8\linewidth]{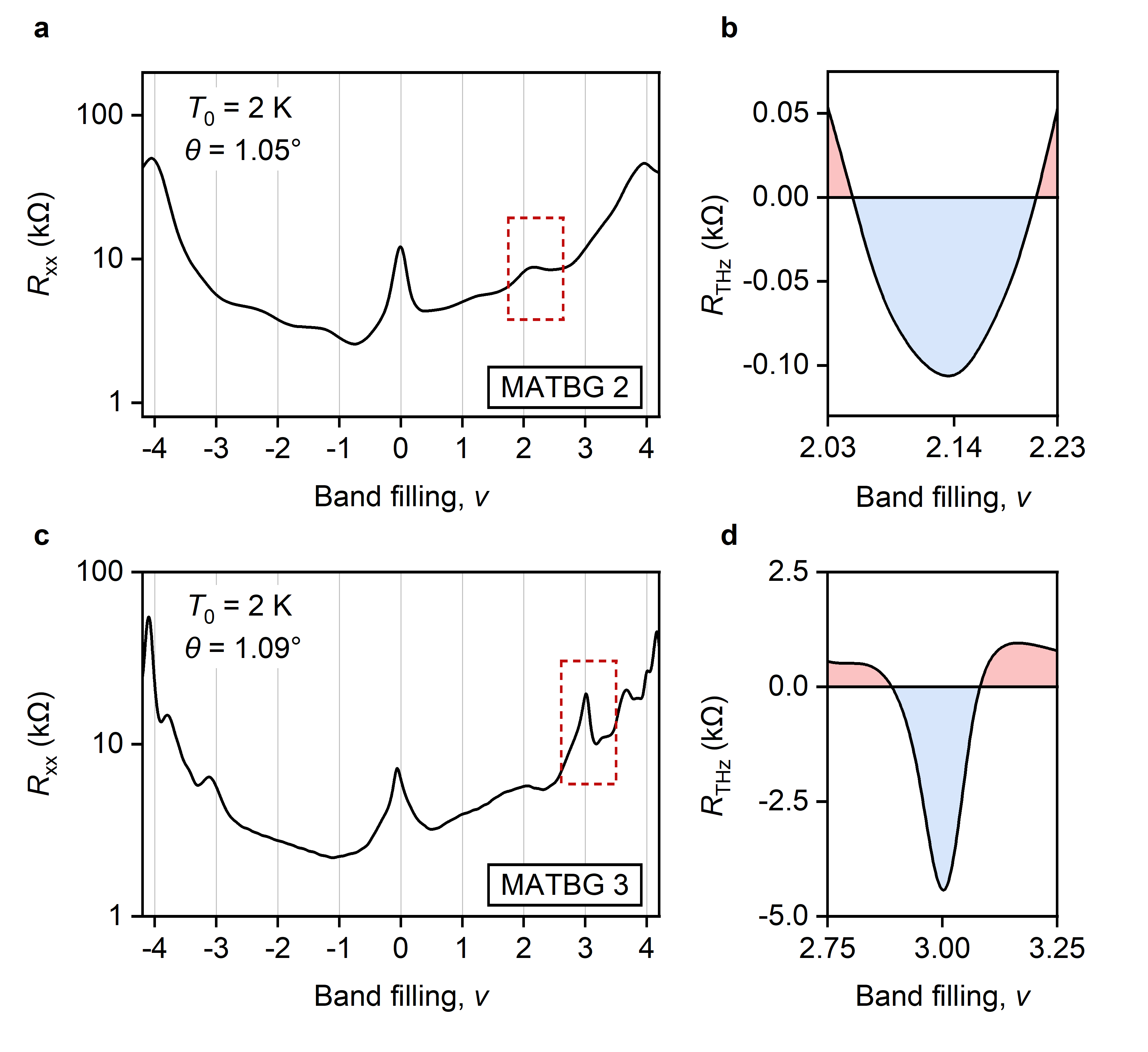}
    \caption{
    \textbf{Photoresistance measurements of different MATBG devices}. \textbf{a}, \textbf{c}, Measurements of the longitudinal resistance $R_\mathrm{xx} (\nu)$ of two additional devices. \textbf{b}, \textbf{d}, Corresponding negative photoresistance at CI features.}
    \label{tbg}
\end{figure}

\newpage
\clearpage
\newpage

\suppnote{NEP, DR and $G_{\mathrm{th}}$ estimation} 
Hot-electron bolometers are generally characterized by Johnson noise ($\mathrm{NEP}_{\mathrm{JN}}$),
thermal fluctuation noise ($\mathrm{NEP}_{\mathrm{TF}}$), and amplifier noise ($\mathrm{NEP}_{\mathrm{amp}}$).
The total noise-equivalent power is therefore given by
\begin{equation}
\mathrm{NEP}_{\mathrm{tot}}
=
\sqrt{
\mathrm{NEP}_{\mathrm{JN}}^{2}
+
\mathrm{NEP}_{\mathrm{TF}}^{2}
+
\mathrm{NEP}_{\mathrm{amp}}^{2}
}.
\end{equation}

Since $\mathrm{NEP}_{\mathrm{amp}}$ depends on the specific measurement scheme,
we consider only the contributions from Johnson noise and thermal fluctuations.
For the Johnson noise estimation, we employ two independent approaches.

The first approach is based on the fundamental expression for
current-biased bolometers~\cite{mather1982bolometer}:
\begin{equation}
\mathrm{NEP}^\mathrm{F}_{\mathrm{JN}}
=
\sqrt{4 k_{\mathrm{B}} T_{\mathrm{e}} P}
\left(
\frac{(\frac{\mathrm{d}V}{\mathrm{d}I})^{*} + V^{*}/I^{*}}{(\frac{\mathrm{d}V}{\mathrm{d}I})^{*} - V^{*}/I^{*}}
\right)^{}
\left(1 + \omega^{2}\tau^{2}\right),
\end{equation}
where $\tau = C/G_{\mathrm{th}}$ is the characteristic thermal relaxation time of the
bolometer with electronic heat capacity $C$ and thermal conductance $G_{\mathrm{th}}$,
$V^{*}$ and $I^{*}$ are the voltage and current at the operating point,
$k_{\mathrm{B}}$ is the Boltzmann constant, and $P$ is the dissipated
DC power at the operating point.
In our experiment, the modulation frequency is much smaller than the thermal
relaxation rate $1/\tau = G_{\mathrm{th}}/C$, such that $\omega\tau \ll 1$.
Using this expression, we obtain
$\mathrm{NEP}^{F}_{\mathrm{JN}} = 5\times10^{-16}\,~\mathrm{W/\sqrt{Hz}}$.

The second method estimates the Johnson noise contribution directly from the ratio
\begin{equation}
\mathrm{NEP}^{\mathrm{exp}}_{\mathrm{JN}} = \frac{S_\mathrm{V}}{R^{\mathrm{}}_\mathrm{V}}\sim2\cdot10^{-11}~\mathrm{W/\sqrt{Hz}},
\end{equation}

where $S_\mathrm{V}=\sqrt{4k_{\mathrm{B}}T_{\mathrm{e}}R_{}}$ is the Johnson voltage noise
spectral density and $R^{\mathrm{}}_{V}$ is the external voltage responsivity. 

The thermal fluctuation noise can be estimated using

\begin{equation}
\mathrm{NEP}_{\mathrm{TF}}
=
\sqrt{
4 k_{\mathrm{B}} T_{\mathrm{e}}^{2}
G_{\mathrm{th}}(T_{\mathrm{e}})
}.
\end{equation}

\begin{figure*}[ht!]
  \centering\includegraphics[width=1\linewidth]{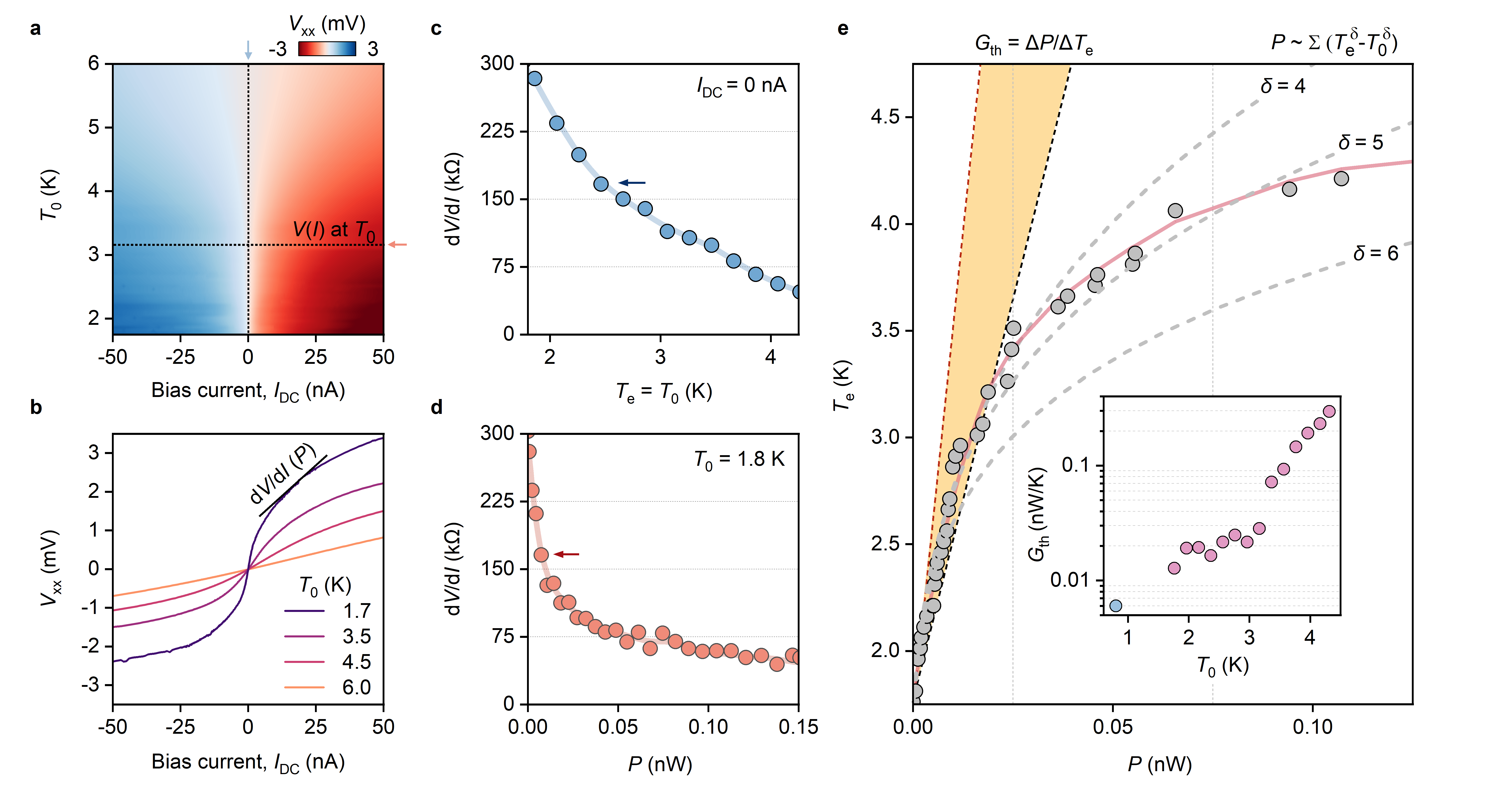}
    \caption{\textbf{Estimation of the thermal conductance.} 
\textbf{a}, Four-probe DC voltage-drop map at $\nu = 2$ as a function of the cryostat base temperature $T_\mathrm{0}$ and DC current $I_\mathrm{DC}$ in the correlated-insulating (CI) state. 
\textbf{b}, $I$–$V$ characteristics at selected cryostat base temperatures $T_\mathrm{0}$. 
\textbf{c}, Differential resistance at $I_\mathrm{DC} = 0$~nA as a function of $T_\mathrm{e}=T_\mathrm{0}$. 
\textbf{d}, Reduction of the differential resistance as a function of the absorbed DC power at $T_\mathrm{0} = 1.8$~K. 
\textbf{e}, Estimation of the electronic temperature rise induced by DC heating. The graph shows the deviation from the expected power-law behavior. Inset: thermal conductance $G_\mathrm{th} = \Delta P / \Delta T_\mathrm{e}$ in the linear regime as a function of $T_\mathrm{0}$. The blue data point corresponds to thermal conductance of the flat band in MATBG reported in Ref.~\cite{supp:di2022revealing}.
}
	\label{thermal_cond}
\end{figure*}

For current-biased bolometers, the thermal conductance $G_\mathrm{th}$ generally depends on temperature; therefore, operating the device at a finite bias current increases the electronic temperature $T_\mathrm{e}$ and shifts $G_\mathrm{th}$ away from its base-temperature value. In principle, $G_\mathrm{th}$ at the operating point can be obtained by measuring the thermal relaxation time $\tau=C/G_\mathrm{th}$, if the heat capacity $C$ is known. Here, instead, we estimate $G_\mathrm{th}$ at the operating point using a calibration based on the temperature dependence of the differential resistance in the correlated-insulator (CI) state.
To determine $G_{\mathrm{th}}$, we assume that in the CI state the resistance is predominantly governed by the electronic temperature $T_\mathrm{e}$. We measured a set of two-terminal $I$--$V$ characteristics at filling factor $\nu=2$ for a range of bath temperatures $T_0$ (Supplementary Figure~\ref{thermal_cond}a) and extracted the zero-bias differential resistance $(\frac{\mathrm{d}V}{\mathrm{d}I})_{I_\mathrm{DC}= 0}$ as a function of temperature, where $T_\mathrm{e}=T_0$ in equilibrium (Supplementary Figure~\ref{thermal_cond}c). In addition, for each base temperature we measured $(\frac{\mathrm{d}V}{\mathrm{d}I})_{I_\mathrm{DC}}$ as a function of the applied DC heating power $P=I\,V_{\mathrm{2pt}}$ (Supplementary Figure~\ref{thermal_cond}d). Thus, by matching the drop in $\frac{\mathrm{d}V}{\mathrm{d}I}$ measured at $T_\mathrm{e}=T_0$ with the drop in $\frac{\mathrm{d}V}{\mathrm{d}I}$ associated with the absorption of DC power $P$, we obtain the calibration curve $T_\mathrm{e}(P)$ shown in Supplementary Figure~\ref{thermal_cond}e.

The resulting nonlinear $T_\mathrm{e}(P)$ dependence suggests a formation of a thermal quasiequilibrium between hot electrons and a cold phonon bath. In this regime, strong thermal decoupling between electrons and the lattice enables efficient energy relaxation via phonon-mediated processes, whose power can be phenomenologically described as 
\begin{equation}\label{Pe-ph}
    P_\mathrm{} = \Sigma_\mathrm{e-ph}A(T_\mathrm{e}^\delta - T_\mathrm{0}^\delta),
\end{equation}
where $\Sigma_\mathrm{e-ph}$ is the effective electron--phonon coupling strength, $A$ is the active device area, and the exponent $\delta$ depends on the microscopic details of electron--phonon interactions, such as dimensionality, disorder, screening, and electron--phonon Umklapp scattering. In our data $\delta\approx5$ is within the range $\sim3$--$5$ of experimentally reported electron--phonon cooling scenarios~\cite{fong2013measurement,fong2012ultrasensitive,supp:aamir2021ultrasensitive,betz2013supercollision,lee2020graphene,mckitterick2016electron,Laitinen2014,ELFATIMY2019497}. Since the latter happens at the ps-scale, our CI bolometers can offer intrinsically ultra-fast response.

In the linear regime, the slope $\Delta P/\Delta T_{\mathrm{e}}$ yields the thermal conductance $G_{\mathrm{th}}$ at $T_\mathrm{e}\rightarrow T_{0}$. Having determined $G_{\mathrm{th}}$ at the lowest temperature $T_{0}$, we estimate its value at elevated electronic temperatures using the scaling relation:
$\Delta P / \Delta T_{\mathrm{e}} \propto T_{\mathrm{e}}^{\delta-1}$,
yielding
\begin{equation}
G_{\mathrm{th}}(T_{\mathrm{e}})
=
G_{\mathrm{th}}(T_{0})
\left(
\frac{T_{\mathrm{e}}}{T_{0}}
\right)^{\delta-1}.
\end{equation}

Using $G_{\mathrm{th}}(T_{0}) = 12\times10^{-12}\,~\mathrm{W/K}$ (determined from the experimental data), $\delta \simeq 5$, and the relevant electronic temperature at the operating point,
we obtain
$G_{\mathrm{th}}(T_{\mathrm{e}})\approx2\times10^{-9}\,~\mathrm{W/K}$,
which results in
$\mathrm{NEP}_{\mathrm{TF}} \sim2\times10^{-15}~\mathrm{W/\sqrt{Hz}}$.

Using the internal (absorbed-power) responsivity $R_\mathrm{V}^\mathrm{abs}$, the 
thermal-fluctuation-limited $\mathrm{NEP}$ translates into an equivalent voltage noise 
at the device output:

\begin{equation}
    S_\mathrm{TF} = R_\mathrm{V}^\mathrm{abs} \times \mathrm{NEP}_\mathrm{TF} 
    \approx 3\times10^{-8}~\mathrm{V}/\sqrt{\mathrm{Hz}}.
\end{equation}

Comparing with Johnson--Nyquist-limited readout noise floor, the 
ratio between the two is
\begin{equation}
    \frac{S_\mathrm{TF}}{S_\mathrm{V}} 
    = \frac{3\times10^{-8}~\mathrm{V}/\sqrt{\mathrm{Hz}}}{2\times10^{-9}~\mathrm{V}/\sqrt{\mathrm{Hz}}}
    \approx 15.
\end{equation}
Since $S_\mathrm{TF}$ is about 15 times (more than an order of magnitude) 
larger than $S_\mathrm{V}$, the thermal fluctuation noise dominates over the readout 
noise. The device is therefore thermally limited, with 
$\mathrm{NEP}_\mathrm{TF}$ setting the relevant noise floor.

We define the dynamic range of the device as
\begin{equation}\label{eq:DR_def}
    \mathrm{DR}
    = \frac{P_\mathrm{sat}}{\mathrm{NEP}\,\sqrt{\Delta f}},
\end{equation}

where $P_\mathrm{sat}$ is the absorbed power at which the response deviates from linearity and $\mathrm{NEP}\,\sqrt{\Delta f}$ is the minimum detectable
power in a $\Delta f = 1$~Hz post-detection band.  All powers are referred to the absorbed plane, so that the coupling efficiency cancels in the ratio.
Since the noise floor is set 
by $\mathrm{NEP}_\mathrm{TF}$, the dynamic range is thermally limited:

\begin{equation}
    \mathrm{DR}
    = \frac{P_\mathrm{sat}}{\mathrm{NEP}_\mathrm{TF}\sqrt{\Delta f}}
    = \frac{100~\mathrm{pW}}{2\times10^{-15}~\mathrm{W}}
    \approx 5.0\times10^{4}.
\end{equation}

\newpage

\begin{table*}[t]

\centering
\caption{\textbf{Power and responsivity estimation for mm-wave and FIR radiation.}}
\label{tab:power_responsivity}
\begin{threeparttable}
\renewcommand{\arraystretch}{1.25}
\begin{tabular}{|l|c|c|}
\hline
\textbf{Parameter} & \textbf{0.14 THz} & \textbf{3.5 THz} \\
\hline
Wavelength, $\lambda$ ($\mu$m) & $2140$ & $85$ \\
\hline
TBG area, $S_\mathrm{TBG}$ ($\mu$m$^2$)
& \multicolumn{2}{c|}{$249$} \\
\hline
Beam area, $S_\mathrm{b}$ ($\mathrm{mm}^2$) & $536$ & $1.6$ \\
\hline
Max power density, $\rho_0$ ($\mu$W/mm$^2$)
& $70$ & $77$ \\
\hline\hline
\multicolumn{3}{|l|}{\textbf{Responsivity in the linear regime}} \\
\hline
\makecell[l]{Per measured absorbed power,\\ $R^\mathrm{abs}_\mathrm{V}$ (V/W)}
& \multicolumn{2}{c|}{$(1.7\pm0.4)\cdot10^{7}$} \\
\hline
Per power density, $R^\mathrm{\rho}_\mathrm{V}$ (V$\mathrm{m^{2}}$/W)
& ($2.1\pm0.5)\cdot10^{-4}$ & $\sim3.2\cdot10^{-6}$ \\
\hline
Per total incident power (this device, ideal focusing), $R^\mathrm{NA=1}_\mathrm{V}$ (V/W)
& $75\pm15$ & $\sim370$ \\
\hline
\multicolumn{3}{|l|}{\textbf{NEP estimation}} \\
\hline
Thermal fluctuations, $\mathrm{NEP}_\mathrm{TF}$ (W/$\sqrt{\mathrm{Hz}}$)
& \multicolumn{2}{c|}{$\sim[1.5, ~2.4]\cdot10^{-15}$} \\
\hline
Ideal focusing (this device), $\mathrm{NEP}^\mathrm{exp}_\mathrm{JN}$ (W/$\sqrt{\mathrm{Hz}}$)
& \multicolumn{2}{c|}{$\sim[2.1, ~3.2]\cdot10^{-11}$} \\
\hline
Fundamental limit, $\mathrm{NEP}^{F}_\mathrm{JN}$ (W/$\sqrt{\mathrm{Hz}}$)
& \multicolumn{2}{c|}{$\sim5\cdot10^{-16}$} \\
\hline
\hline
\multicolumn{3}{|l|}{\textbf{Response time estimation}} \\
\hline
RC timescale (this device), $\tau_\mathrm{RC}$ (s)
& \multicolumn{2}{c|}{$\sim50\cdot10^{-9}$} \\
\hline
Intrinsic cooling time \cite{supp:mehew2024ultrafast}, $\tau_\mathrm{}$ (s)
& \multicolumn{2}{c|}{$\sim10^{-12}$} \\
\hline
\end{tabular}
\end{threeparttable}
\vspace{2mm}
\end{table*}

\newpage
\clearpage

\suppnote{THz-induced electron heating and heat transfer balance}
THz absorption deposits power $P_\mathrm{abs}$ into the electronic subsystem of the TBG channel and thereby heats the charge carriers. Direct excitation of optical phonons is strongly suppressed in graphene-based heterostructures because the relevant phonon modes are nonpolar. In addition, optical phonons in the adjacent hBN layers have characteristic energies several orders of magnitude larger than the incident THz photon energy. We therefore attribute the absorbed THz power primarily to selective heating of the electronic system.

In graphene-based structures, inter-carrier thermalization occurs on femtosecond timescales, far shorter than any relevant energy-relaxation process\cite{massicotte2021hot}. The electronic subsystem can therefore rapidly establish a Fermi--Dirac distribution characterized by an effective temperature $T_\mathrm{e}$, while remaining out of equilibrium with the lattice at temperature $T_\mathrm{L}$.

\begin{figure*}[ht!]
\centering\includegraphics[width=1\linewidth]{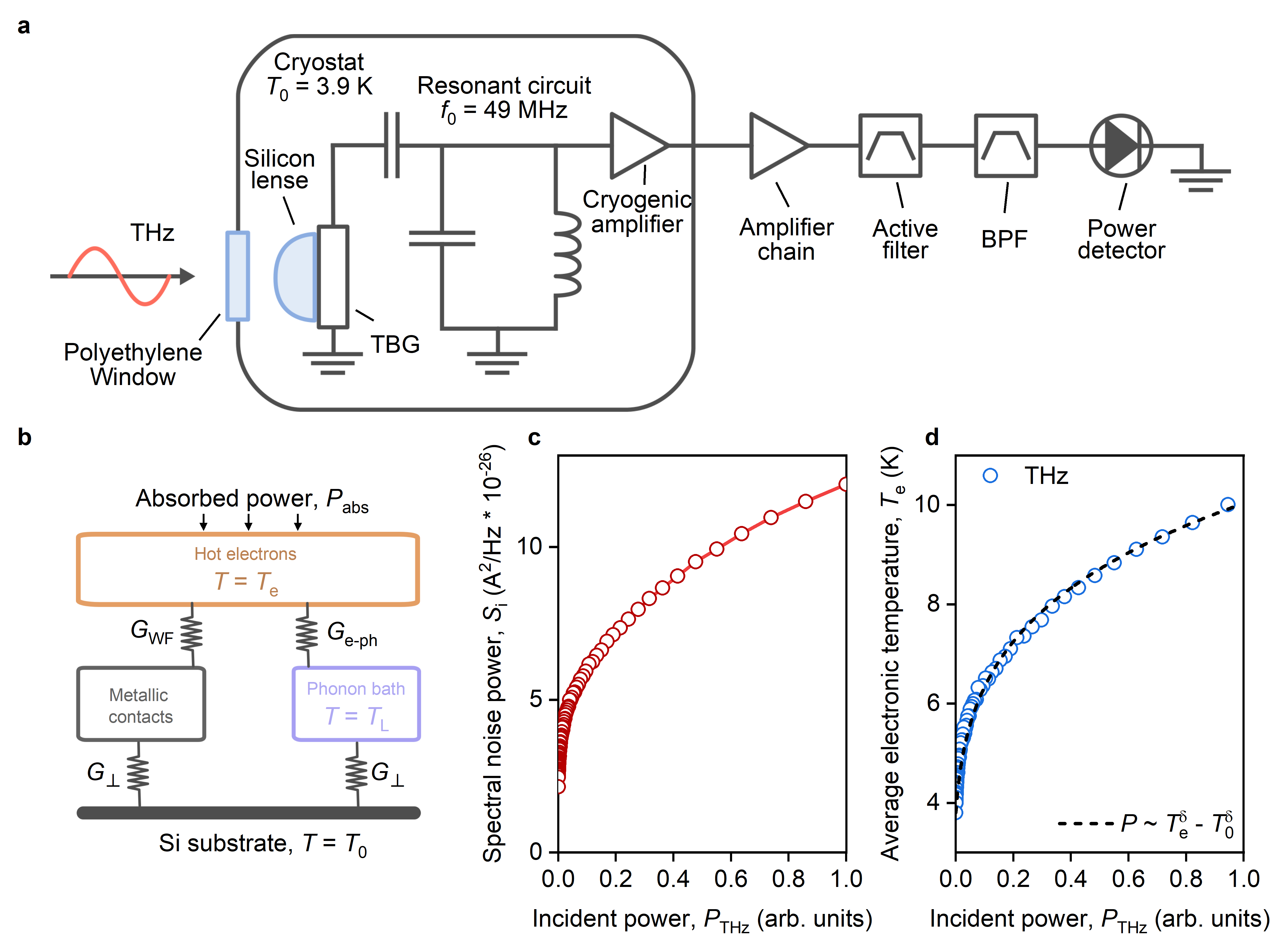}
\caption{\textbf{Electron heating and Johnson noise thermometry.} 
\textbf{a}, Schematic of the noise thermometry measurement setup. 
\textbf{b}, Schematic diagram of heat transfer channels in TBG.  
\textbf{c}, Spectral noise power $S_\mathrm{i}$ measured as a function of the normalized incident THz power, $P_\mathrm{THz}$.
\textbf{d}, Electronic temperature $T_\mathrm{e}$ inferred from the noise measurements in \textbf{c}. The black dashed curve is a fit to the steady-state heat-balance relation $P \propto T_\mathrm{e}^{\delta}-T_0^{\delta}$, yielding $\delta \approx 5$.}
\label{FigS8}
\end{figure*}

Hot-electron relaxation in graphene and related systems has been widely investigated~\cite{fong2012ultrasensitive,massicotte2021hot,fong2013measurement,supp:aamir2021ultrasensitive,lee2020graphene,fried2024performance,efetov2018fast,mckitterick2016electron,mckitterick2013performance,yan2012dual,lara2019towards,shein2024fundamental}. Under continuous excitation, the steady-state value of $T_\mathrm{e}$ is fixed by the balance between absorbed power and electronic cooling. Two heat-removal pathways are relevant. The first is Wiedemann--Franz (WF) diffusion of electronic heat into the cold metallic contacts, described by $P_\mathrm{WF}(T_\mathrm{e})\propto G_\mathrm{WF}$, where $G_\mathrm{WF}$ is the electronic thermal conductance. The second is the emission of low-energy acoustic TBG phonons through disorder- or Umklapp-assisted processes~\cite{supp:mehew2024ultrafast}. This contribution can be written as
\begin{equation}
P_\mathrm{e-ph}(T_\mathrm{e},T_\mathrm{L})
=
G_\mathrm{e-ph}(T_\mathrm{e}-T_\mathrm{L})
\propto
\Sigma_\mathrm{e-ph},
\end{equation}
where $\Sigma_\mathrm{e-ph}$ is the corresponding electron--phonon coupling constant (Supplementary Figure~\ref{FigS8}b). The electronic steady state is thus governed by

\begin{equation}\label{e equil}
    P_\mathrm{abs}=P_\mathrm{WF}+P_\mathrm{e-ph}.
\end{equation}

The power $P_\mathrm{e-ph}$ delivered to the TBG phonon system is subsequently removed through the heterostructure into the Si/SiO$_2$ substrate, which remains thermalized at the cryostat base temperature $T_0$ (Supplementary Figure~\ref{FigS8}b). We describe this vertical heat flow as
\begin{equation}
P_\perp(T_\mathrm{L},T_0)
=
G_\perp(T_\mathrm{L}-T_0),
\end{equation}
where $G_\perp$ is the effective out-of-plane thermal conductance. This quantity includes heat transport through the underlying layers together with Kapitza resistances at the material interfaces. The corresponding lattice heat-balance condition is

\begin{equation}\label{ph equil}
    P_\perp=P_\mathrm{e-ph}\leq P_\mathrm{abs}.
\end{equation}

Equation~\eqref{ph equil} sets an upper limit on the TBG lattice-temperature increase for a given $P_\mathrm{abs}$:
\begin{equation}
    T_{\mathrm{L,max}}-T_0=\frac{P_\mathrm{abs}}{G_\perp}.
\end{equation}
The lattice overheating is therefore controlled only by the efficiency of vertical heat removal and does not depend explicitly on either $T_\mathrm{e}$ or the electron--phonon coupling constant $\Sigma_\mathrm{e-ph}$. By contrast, the steady-state electronic temperature is directly sensitive to $\Sigma_\mathrm{e-ph}$ and to the relative strength of the available cooling channels. It can consequently depend on the device geometry, the distance to metallic contacts, and the competition between WF diffusion and electron--phonon relaxation.

In general, the degree of electronic decoupling from the lattice, $T_\mathrm{e}-T_\mathrm{L}$, is determined by the combined action of $G_\mathrm{WF}$, $\Sigma_\mathrm{e-ph}$, and $G_\perp$. In graphene-based heterostructures, however, the electron--phonon energy-transfer rate is exceptionally weak. The resulting hierarchy
\[
G_\mathrm{e-ph}\ll G_\perp
\]
makes energy transfer from the electronic system to the lattice the principal bottleneck in the thermal pathway. Consequently, over the range of $P_\mathrm{abs}$ relevant to our measurements, the electronic and lattice temperature rises satisfy

\begin{equation}\label{te-tl}
    T_\mathrm{e}-T_\mathrm{L}
    \gg
    T_\mathrm{L}-T_0,
\end{equation}

while the TBG lattice remains close to the cryostat temperature,

\begin{equation}\label{tl-t0}
    T_\mathrm{L}-T_0\ll T_0.
\end{equation}

We therefore use the approximation $T_\mathrm{L}=T_0$ throughout the analysis and express the electronic overheating as $T_\mathrm{e}-T_0$. The validity of Eqs.~\eqref{te-tl} and \eqref{tl-t0} under our experimental conditions is established independently by the noise-thermometry and thermal-management measurements discussed below.

\newpage
\suppnote{Noise thermometry in TBG} 
We independently quantified the electronic heating produced by THz radiation using noise thermometry in an additional TBG device. The measurements were carried out in a Scontel closed-cycle cryostat with a base temperature of $T_0=3.9$~K. Optical access was provided through a polyethylene window combined with a Zitex infrared filter, as illustrated in Supplementary Figure~\ref{FigS8}a. The transmitted CW THz radiation was focused onto the device using a silicon lens. The average electronic temperature was determined from the current-noise spectral density $S_\mathrm{i}$ following established noise-thermometry protocols~\cite{PhysRevB.90.161405,Piatrusha2018}. For noise thermometry, the current-noise spectral density $S_\mathrm{i}$ was detected using a resonant tank circuit at the input of a custom-built low-noise amplifier. The amplified signal was further boosted by a low-noise amplifier chain, filtered by band-pass filters, and recorded with a power detector. A detailed description of the detection setup is provided in Ref.~\cite{Noise}. As the incident THz power is increased, the measured $S_\mathrm{i}$ rises substantially, demonstrating pronounced heating of the electronic system under irradiation (Supplementary Figure~\ref{FigS8}c). Conversion of the noise signal into $T_\mathrm{e}$ reveals a strongly sublinear power dependence: the electronic temperature grows rapidly in the weak-excitation regime and evolves much more slowly at higher powers, approaching saturation (Supplementary Figure~\ref{FigS8}d). Such behavior is expected when the deposited energy is removed predominantly through electron--phonon relaxation.

\newpage
\clearpage

\suppnote{Lattice temperature}
To estimate lattice overheating in our experiments, we studied a reference TBG device with a twist angle $\theta \approx 1\degree$. The device incorporates superconducting aluminum contacts, which limit Wiedemann--Franz (WF) heat transport through the leads and thereby favor heat removal through the substrate, as well as a bottom graphite gate and a nearby top graphite heater contacted by two aluminum leads. This geometry enables us to compare two distinct heating configurations: \textbf{(i)} Joule heating produced by a DC current flowing directly through the TBG channel and \textbf{(ii)} heating generated in the graphite layer and delivered vertically through the heterostructure. These configurations differ in how the electronic temperature $T_\mathrm{e}$ is established. For direct current heating, which closely resembles THz excitation, the electronic system can become substantially hotter than the lattice, and its steady-state temperature is controlled primarily by energy relaxation through acoustic-phonon emission. In contrast, heating from the graphite layer raises the temperature of the heterostructure as a whole, such that the electrons and lattice remain approximately equilibrated, $T_\mathrm{e}=T_\mathrm{L}$, with their temperature governed by the cross-plane thermal conductance $G_\perp$.

The reference device exhibits a strong temperature dependence of the longitudinal resistance $R_\mathrm{xx}$ throughout the flat-band doping range, as shown in Supplementary Figure~\ref{FigS9}a. Provided that the steep insulating-like variation of $R_\mathrm{xx}(T_0)$ is primarily electronic in origin, the resistance can serve as a thermometer for the electronic system. We therefore use it to extract $T_\mathrm{e}$ as a function of the power deposited either directly in the TBG channel, $P_\mathrm{dc}$, or in the graphite heater, $P_\mathrm{heater}$. The measurements described below were performed at filling factor $\nu=2.6$.

For the direct-heating calibration, a small AC excitation current, $I_\mathrm{ac}=50$~pA, was superimposed on a larger DC current $I_\mathrm{dc}\sim10$~nA. The four-terminal longitudinal resistance $R_\mathrm{xx}$ and the two-terminal differential resistance $R_\mathrm{2pt}$ were recorded concurrently. The Joule power released in the TBG channel was evaluated from
\begin{equation}
    P_\mathrm{dc}(I_\mathrm{dc})
    =
    I_\mathrm{dc}
    \int\limits_0^{I_\mathrm{dc}}
    R_\mathrm{2pt}(j)\,\mathrm{d}j.
\end{equation}
The corresponding electronic temperature increase, $\Delta T_\mathrm{e}(P_\mathrm{dc})$, was then determined by comparing the measured $R_\mathrm{xx}(P_\mathrm{dc})$ with the independently calibrated temperature dependence $R_\mathrm{xx}(T_0)$, as illustrated in Supplementary Figure~\ref{FigS9}c.

In the second configuration, the TBG channel was kept unbiased, while a current $I_\mathrm{heater}$ was passed through the graphite heater. The voltage across the heater, $V_\mathrm{heater}$, was measured simultaneously, giving the total heater power
\begin{equation}
    P_\mathrm{heater}
    =
    I_\mathrm{heater}V_\mathrm{heater}.
\end{equation}
Because the heater is contacted by superconducting aluminum leads, Joule dissipation in the leads is strongly suppressed, and the applied power is released predominantly within the graphite heater before flowing vertically through the device stack into the substrate. The temperature rise $\Delta T_\mathrm{e}(P_\mathrm{heater})$ was extracted using the same resistance-thermometry procedure, by mapping $R_\mathrm{xx}(P_\mathrm{heater})$ onto the calibrated $R_\mathrm{xx}(T_0)$ curve (Supplementary Figure~\ref{FigS9}d).

Supplementary Figure~\ref{FigS9}e compares the values of $P_\mathrm{dc}$ and $P_\mathrm{heater}$ required to produce the same increase in electronic temperature. The two powers differ by approximately four orders of magnitude. Within this comparison,
\begin{equation}
    \frac{P_\mathrm{dc}}{P_\mathrm{heater}}
    \approx
    \frac{G_\mathrm{e-ph}}{G_\perp},
\end{equation}
where $G_\mathrm{e-ph}$ characterizes energy transfer from electrons to acoustic phonons and $G_\perp$ describes the subsequent cross-plane heat flow into the substrate. The measurements therefore give the order-of-magnitude estimate
\begin{equation}
    \frac{G_\mathrm{e-ph}}{G_\perp}\sim10^{-4}.
\end{equation}
Furthermore, suppressing superconductivity in the aluminum leads with a magnetic field produces no detectable change in the electronic overheating measured in either configuration (Supplementary Figure~\ref{FigS9}c,d). This insensitivity confirms that WF-mediated heat transport through the contacts does not appreciably contribute to cooling of the TBG device at cryogenic temperatures.

Under THz irradiation strong enough to melt the correlated-insulating state in MATBG, the electronic temperature typically increases by $\Delta T_\mathrm{e}\sim10$~K. Combining this value with $G_\mathrm{e-ph}/G_\perp\sim10^{-4}$ and accounting for the nonlinear saturation of $T_\mathrm{e}(P_\mathrm{abs})$, we obtain a conservative upper estimate for the associated lattice-temperature increase of
\begin{equation}
    \Delta T_{\mathrm{L,max}}\sim100~\mathrm{mK}.
\end{equation}

\begin{figure*}[ht!]
  \centering\includegraphics[width=1\linewidth]{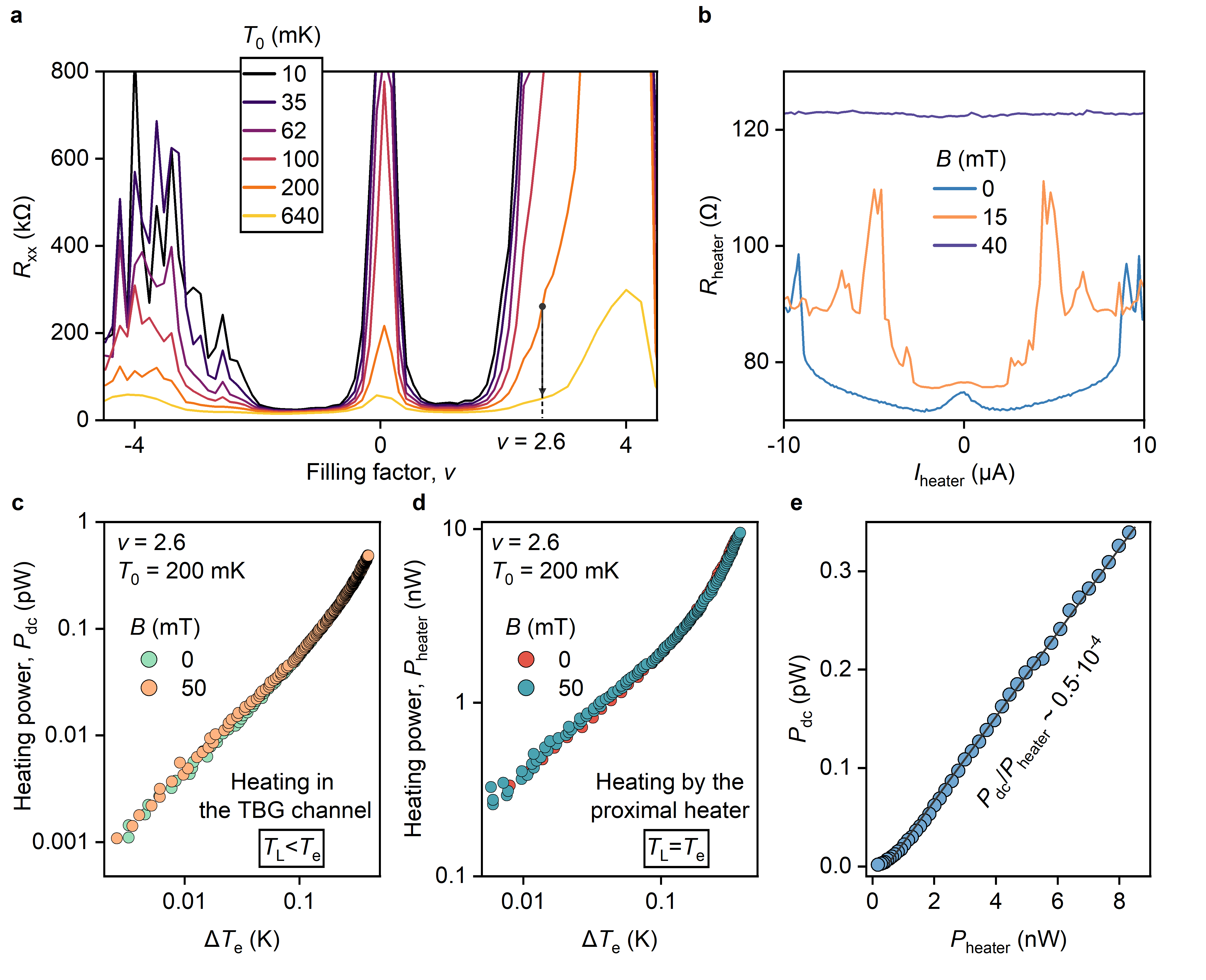}
    \caption{\textbf{Electron heating and thermalization in TBG at $T_0 = 200~\mathrm{mK}$.} \textbf{a}, Longitudinal resistance of the $\theta\approx 1\degree$ device plotted against the moir\'e band filling $\nu$ for several representative bath temperatures $T_0$. The excitation current was kept below $I_\mathrm{ac} = 50\;$pA so that the electronic system remained in thermal equilibrium with the lattice. \textbf{b}, Heater resistance $R_\mathrm{heater}$ versus the applied DC current $I_\mathrm{heater}$, recorded at $B = 0$, 15 and 40\;mT ($T_0 = 200\;$mK). \textbf{c}, DC power $P_\mathrm{dc}$ as a function of the electron temperature rise $\Delta T_\mathrm{e}$, extracted from the condition $R_\mathrm{xx}(T_0) = R_\mathrm{xx}(P_\mathrm{dc})$ at $\nu = 2.6$. \textbf{d}, Analogous calibration performed with the heater: $P_\mathrm{heater}$ versus $\Delta T_\mathrm{e}$, obtained by equating $R_\mathrm{xx}(T_0)$ and $R_\mathrm{xx}(P_\mathrm{heater})$ at $\nu = 2.6$. \textbf{e}, Values of $P_\mathrm{dc}$ and $P_\mathrm{heater}$ that produce an identical rise in $T_\mathrm{e}$ at $\nu = 2.6$, plotted against each other.
}
	\label{FigS9}
\end{figure*}

\newpage
\clearpage

\suppnote{Electromagnetic simulations}

\begin{figure}[ht!]
\centering
\begin{minipage}[c]{0.57\linewidth}
  \centering
  \includegraphics[width=\linewidth]{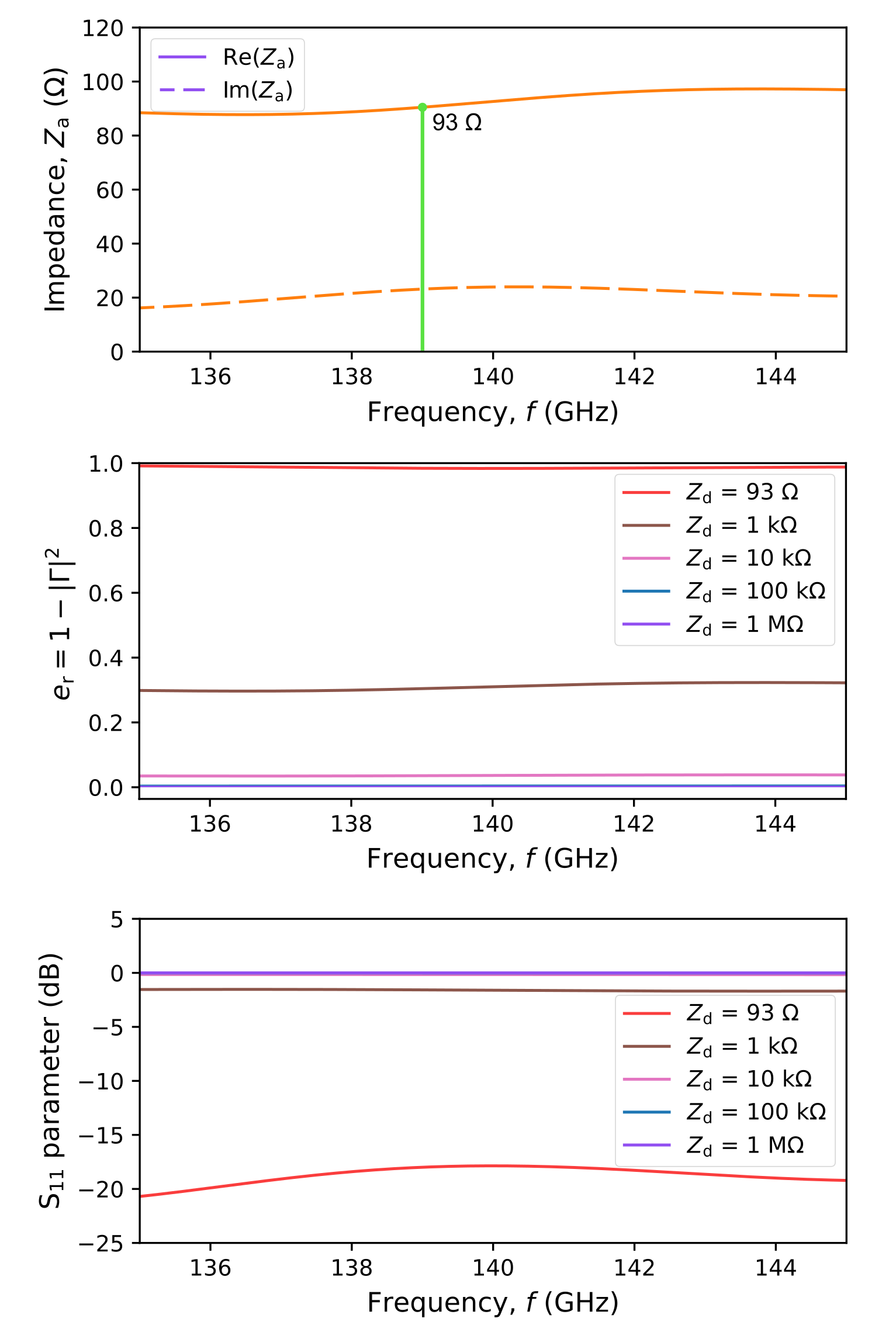}
\end{minipage}\hfill
\begin{minipage}[c]{0.39\linewidth}
  \caption{\textbf{Antenna impedance, absorption, and reflection in the 135--145 GHz band.}
  \textbf{a}, Real and imaginary parts of the antenna input impedance $Z_\mathrm{a}$ as a function of frequency, showing a predominantly resistive response with $Z_\mathrm{a} \approx 93~\mathrm{\Omega}$ near $140$~GHz. \textbf{b}, Simulated absorption for five detector load impedances: a matched reference load ($Z_\mathrm{d} = 93~\mathrm{\Omega}$) and four high-impedance loads representative of the device ($Z_\mathrm{d} = 1~\mathrm{M\Omega};~ 100~\mathrm{k\Omega};~ 10~\mathrm{k\Omega};~ 1~\mathrm{k\Omega} $). \textbf{c}, Corresponding reflection coefficient $S_{11}$ for different load conditions.}
  \label{simulation}
\end{minipage}
\end{figure}

To understand why the antenna does not contribute to the bolometric response, we performed electromagnetic simulations of the full device geometry. The simulations were performed using the frequency-domain solver in CST Studio Suite. The analysis was restricted to the 135--145~GHz band in order to accurately resolve the antenna impedance and coupling behavior in the frequency range of interest. The bow-tie antenna was excited through a localized feed region, where the detector was represented by an effective lumped impedance. Five load conditions were considered: a matched reference load of $Z_\mathrm{d} = 93~\mathrm{\Omega}$, and a high-impedance load of $Z_\mathrm{d} = 1~\mathrm{M\Omega};~ 100~\mathrm{k\Omega};~ 10~\mathrm{k\Omega};~ 1~\mathrm{k\Omega} $. To ensure accurate resolution of the near-field distribution and current flow in the feed region, a local mesh refinement was applied. The maximum mesh step in this region was constrained to $\Delta \le \frac{\lambda_{\min}}{15}$,
where $\lambda_{\min}$ corresponds to the wavelength at the upper edge of the simulated band ($145$~GHz) in the surrounding medium. This refinement was applied only in the vicinity of the feed and antenna gap in order to limit the total computational cost. 
To directly reflect the power delivered to the device in
the experiment, we report the power transfer efficiency $e_r = 1 - |\Gamma|^2$ -- the
fraction of the antenna's available power delivered into the device. This is
$e_r \approx 4\times10^{-4}$ at $Z_\mathrm{d} = 1~\text{M}\mathrm{\Omega}$ (dark),
$4\times10^{-3}$ at $100~\text{k}\mathrm{\Omega}$, $0.04$ at $10~\text{k}\mathrm{\Omega}$, and $0.30$ at
$1~\text{k}\mathrm{\Omega}$, versus $e_r \approx 0.98$ for the matched case $Z_\mathrm{d} = 93~\mathrm{\Omega}$.
Across the experimentally relevant illuminated range ($Z_\mathrm{d} \gtrsim 10~\text{k}\mathrm{\Omega}$)
the antenna-mediated power transfer is suppressed more than 25-fold relative to the
matched case and stays at the few-percent level or below. This analysis therefore indicates that the antenna-coupled channel constitutes only a minor contribution to power absorption under illumination. This conclusion is consistent with the observed polarization-independent photoresistance, which points to direct free-space absorption, rather than coupling through the polarization-selective antenna, as the dominant mechanism.

\newpage

\suppnote{Correlated insulator vs superconducting bolometry}

\begin{figure}[ht!]
    \centering
    \includegraphics[width=1\linewidth]{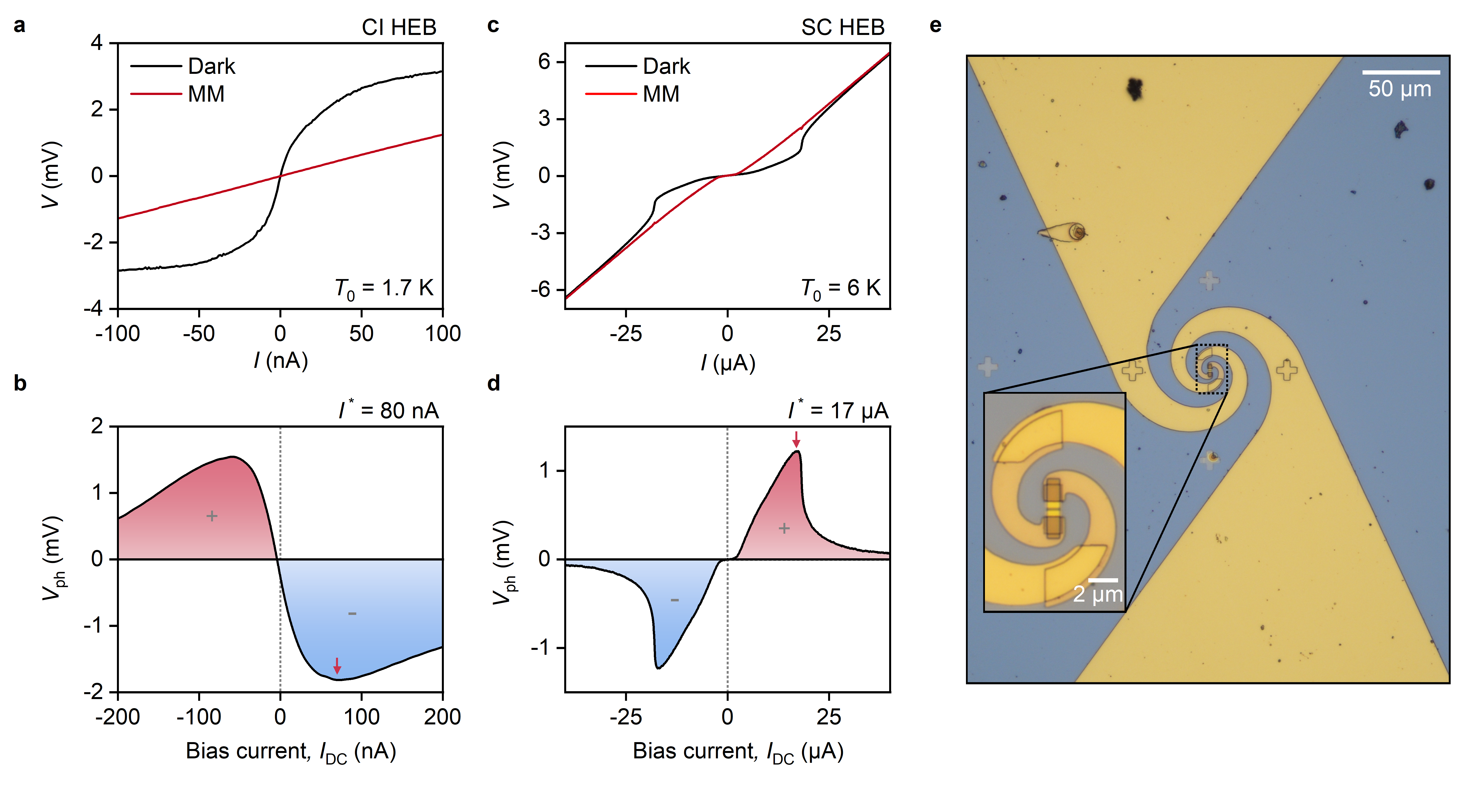}
    \caption{\textbf{Comparison of the CI and SC bolometry.}
\textbf{a}, $V$--$I$ characteristics of the Correlated Insulator HEB measured in the dark and under MM-wave irradiation.
\textbf{b}, Corresponding photovoltage of the CI HEB as a function of the DC bias current $I_\mathrm{DC}$.
\textbf{c}, $V$–-$I$ characteristics of a NbN SC HEB measured in the dark and under MM-wave irradiation.
\textbf{d}, Corresponding photovoltage of the NbN SC HEB as a function of the DC bias current $I_\mathrm{DC}$.
\textbf{e}, Optical micrograph of the NbN SC HEB integrated with a spiral antenna. }
    \label{SCHEB}
\end{figure}

Supplementary Figure~\ref{SCHEB} presents a comparison of two different bolometric principles based on a CI and superconductivity. Supplementary Figure~\ref{SCHEB}a,b show the $I$--$V$ characteristics and the corresponding THz-induced photovoltage of the CI hot-electron bolometer (HEB) measured at $T = 1.7$~K. Supplementary Figures~\ref{SCHEB}c,d present analogous measurements for a commercial NbN HEB measured near the critical temperature. The detection mechanisms in the two devices are effectively symmetric. In the CI, THz-induced heating suppresses the CI state and leads to a decrease in resistance, whereas in the SC HEB, heating weakens superconductivity and causes an increase in resistance. As a result, the photovoltage signals in the two devices have opposite signs under MM-wave excitation. Interestingly, an important distinction lies in the required DC bias current $I^*$ at the operating point. The superconducting HEB operates at a much higher bias current ($\sim 17~\mu$A) compared to the CI-based detector ($\sim 80$~nA). 

\textbf{Comparison of dynamic ranges.} For a representative phonon-cooled NbN HEB
($T \approx 9$~K, normal-state resistance $R \approx 50~\Omega$, thermal conductance $G \approx 16~\mathrm{nW/K}$, transition width $\Delta T_\mathrm{c} \approx 1$~K, direct-detection responsivity $R_\mathrm{V} \approx 2\times10^{4}$~V/W~\cite{seliverstov2014fast}) we obtain
\begin{equation}
    S_\mathrm{V} = \sqrt{4 k_\mathrm{B} T R}
    \approx 1.6\times10^{-10}~\mathrm{V}/\sqrt{\mathrm{Hz}},
\end{equation}
\begin{equation}
    \mathrm{NEP}_\mathrm{TF}
    = \sqrt{4 k_\mathrm{B} T^2 G}
    \approx 8.5\times10^{-15}~\mathrm{W}/\sqrt{\mathrm{Hz}},
    \qquad
    \mathrm{NEP}_\mathrm{JN}
    = \frac{S_\mathrm{V}}{R_\mathrm{V}}
    \approx 7.9\times10^{-15}~\mathrm{W}/\sqrt{\mathrm{Hz}}.
\end{equation}

The two contributions are comparable ($S_\mathrm{TF}/S_\mathrm{V} \approx 1$): in contrast to our thermally-limited device, the HEB is co-limited by thermal-fluctuation and Johnson (readout) noise,
$\mathrm{NEP}_\mathrm{tot} \approx 1.2\times10^{-14}~\mathrm{W}/\sqrt{\mathrm{Hz}}$.
The response saturates when the absorbed power sweeps the electron temperature across the full transition.
 
Combining these for $P_\mathrm{sat} \approx0.5G\Delta T_\mathrm{c} \approx 8$~nW,
\begin{equation}
    \mathrm{DR}_\mathrm{HEB}
    = \frac{P_\mathrm{sat}}{\mathrm{NEP}_\mathrm{tot}\sqrt{\Delta f}}
    \approx 7\times10^{5}~\,
\end{equation}

Thus, the bandwidth-normalized dynamic range of a representative superconducting NbN HEB is approximately one order of magnitude larger than that of our CI HEB. We emphasize, however, that this comparison corresponds to the conventional 1-Hz noise bandwidth, for which the lower detectable power is set by the NEP. In an actual measurement with bandwidth $\Delta f$, the noise-equivalent input power increases as $\mathrm{NEP}\sqrt{\Delta f}$, and the usable dynamic range is therefore reduced as $\frac{1}{\sqrt{\Delta f}}$. The values quoted above should therefore be understood as a common, bandwidth-normalized figure of merit rather than the dynamic range available in a practical finite-bandwidth readout.

\textbf{Comparison of the operating points.} In direct detection, the operating point of an HEB is set by the total thermal balance. The bath is held close to $T_\mathrm{c}$, while the DC bias provides Joule heating, so that the electronic temperature $T_\mathrm{e}$ lies on the steepest part of the superconducting transition, where $|\mathrm{d}R\mathrm{/d}T_\mathrm{e}|$ and the responsivity are maximal. Absorbed background radiation enters the same thermal balance as an additional heating term. A change in absorbed power therefore changes the mean electronic temperature by \begin{equation} \delta T_\mathrm{e} = \frac{\delta P_\mathrm{abs}}{G}, \end{equation} which moves the device along the transition. Thus, the radiation background not only produces a signal, but also shifts the operating point itself; unless the bias or bath temperature is readjusted, the detector is displaced from the optimal point. Therefore, direct-detection operation generally requires re-biasing when the incident power changes. Consequently, the apparently large formal dynamic range of an HEB is obtained by treating $V_\mathrm{ph}(P_\mathrm{abs})$ as a curve in which each value of $P_\mathrm{abs}$ defines a new operating point on the nonlinear $I$--$V$ characteristic. As a result, this formal dynamic range does not correspond to the range accessible in a single stable bias configuration.

\begin{figure}[ht!]
\centering
\includegraphics[width=0.45\linewidth]{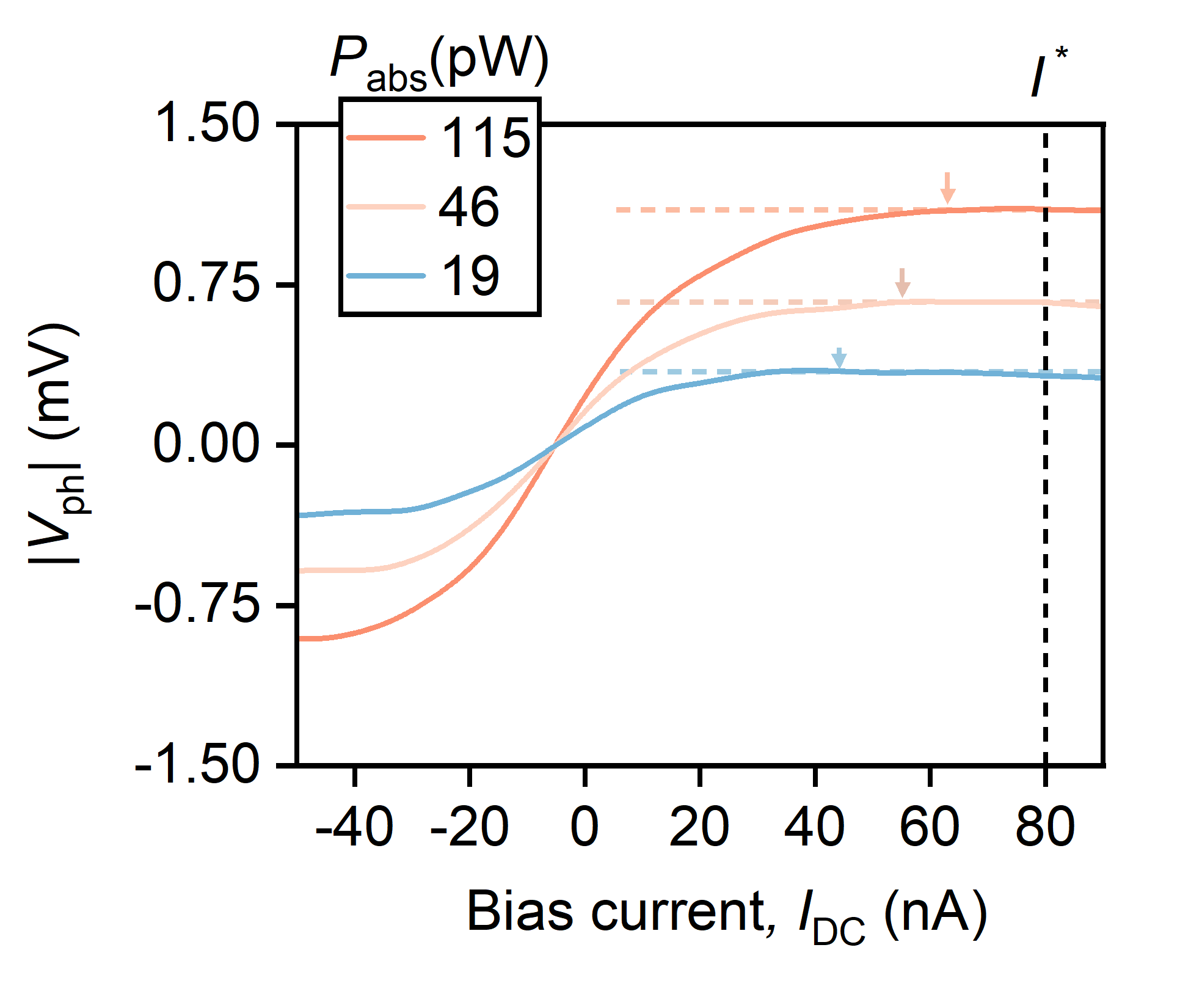}
\caption{
\textbf{Operating point of CI bolometer.}
Photovoltage \(V_{\mathrm{ph}}\) as a function of DC current bias for selected absorbed powers of the incident radiation. At lower radiation intensity, the response reaches its maximum at lower bias currents. Therefore, choosing the working point based on the maximum irradiation power does not reduce the responsivity at lower powers.
}
\label{working}
\end{figure}

Although the CI bolometer requires a DC current bias for readout as well, this bias does not have to be continuously adjusted to maintain a high responsivity. The operating point can be fixed once, for example using the response at the highest incident power. As the incident power density is reduced, the absolute photoresistance decreases and the plateau in \(V_{\mathrm{ph}}(I_{\mathrm{DC}})\) is reached at lower bias currents (see Supplementary Figure~\ref{working}). Thus, a current chosen to place the device in a nearly bias-independent regime at the maximum incident power also keeps it in this regime at lower powers. The full operational dynamic range can therefore be accessed without re-biasing. We note, however, that a small re-optimization of the bias current may still be beneficial for minimizing the thermal-fluctuation noise, $\mathrm{NEP}_{\mathrm{TF}}$, because the DC bias sets the dark-state electronic temperature at the operating point.

\newpage

\suppnote{CI phase diagram}

\begin{figure}[htbp]
    \centering
    \includegraphics[width=0.35\linewidth]{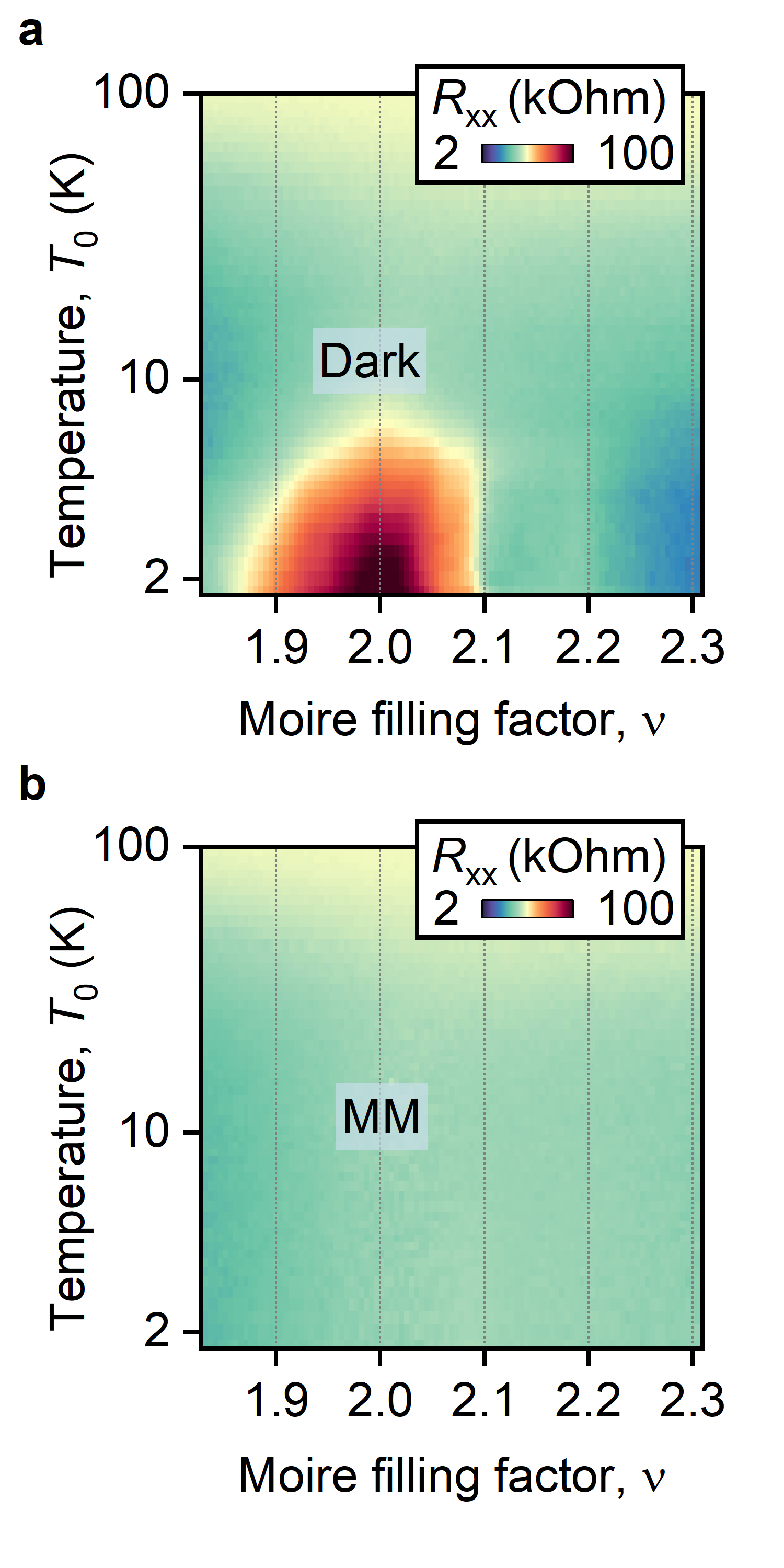}
   \caption{\textbf{Phase diagram of MATBG in the dark and under MM-wave exposure.} \textbf{a-b}, Maps of longitudinal resistance $R_\mathrm{xx}(\nu,T)$ in the dark (a) and under MM-wave illumination (b). }
    \label{fig:discussion}
\end{figure}

\newpage
\clearpage
\section*{Supplementary References}

\begingroup
\let\MergedSupplementaryBibLabel\label
\renewcommand{\label}[1]{\MergedSupplementaryBibLabel{suppbib:#1}}

\endgroup

\end{document}